\pdfoutput=1
\documentclass[aps,twocolumn,showpacs,showkeys,superscriptaddress,amsmath,
amssymb,nofootinbib,floatfix,prc]{revtex4-1}

\usepackage{graphicx}

\usepackage{amsmath}

\usepackage{color}

\def\x{{\boldsymbol x}}
\def\b{{\boldsymbol b}}
\def\y{{\boldsymbol y}}
\def\r{{\boldsymbol r}}
\def\R{{\boldsymbol R}}
\def\z{{\boldsymbol z}}
\def\s{{\boldsymbol s}}
\def\T{{\boldsymbol T}}
\def\D{{\boldsymbol \Delta}}
\newcommand{\rmd}{{d}}
\newcommand{\beq}{\begin{eqnarray}}
\newcommand{\eeq}{\end{eqnarray}}

\makeatletter

\begin{document}

\title{Correlations in the Monte Carlo Glauber model}

\author{Jean-Paul Blaizot}
\email{Jean-Paul.Blaizot@cea.fr} 
\affiliation{Institut de Physique Th\'eorique, CNRS/URA 2306, F-91191 Gif-sur-Yvette, France}
\author{Wojciech Broniowski}
\email{Wojciech.Broniowski@ifj.edu.pl} 
\affiliation{Institut de Physique Th\'eorique, CNRS/URA 2306, F-91191 Gif-sur-Yvette, France}
\affiliation{The H. Niewodnicza\'nski Institute of Nuclear Physics, Polish Academy of Sciences, PL-31342 Krak\'ow, Poland} 
\affiliation{Institute of Physics, Jan Kochanowski University, PL-25406~Kielce, Poland}
\author{Jean-Yves Ollitrault}
\email{Jean-Yves.Ollitrault@cea.fr} 
\affiliation{Institut de Physique Th\'eorique, CNRS/URA 2306,
F-91191 Gif-sur-Yvette, France}

\date{14 May 2014}

\begin{abstract}
Event-by-event fluctuations of observables are often modeled using the
Monte Carlo Glauber model, in which the energy is initially deposited
in sources associated with wounded nucleons. In this paper, we analyze
in detail the correlations between these sources in proton-nucleus
and nucleus-nucleus collisions. 
There are correlations arising from nucleon-nucleon correlations
within each nucleus, and correlations due to the collision mechanism,
which we dub twin correlations. We investigate this new 
phenomenon in detail. At the RHIC and LHC energies, correlations are found to
have 
modest effects on size and eccentricity fluctuations, such that the
Glauber model produces to a good approximation a collection of
independent sources. 
\end{abstract}



\maketitle

\section{Introduction}
Quantum fluctuations in the wave functions of colliding nuclei at
ultrarelativistic energies result in energy-density correlations in the initial
stage of these collisions. These density correlations in turn produce 
observable correlations, which have been actively studied over the past few 
years~\cite{Adare:2012kf}. 
There are two sources of such quantum fluctuations: fluctuations of
positions of nucleons within the nucleus, and fluctuations at the
subnucleonic level. 
In this paper, we analyze the first source, and the resulting density
correlations in the proton-nucleus and nucleus-nucleus collisions, using
the Glauber approach~\cite{Glauber:1959aa,Czyz:1969jg,Miller:2007ri,Broniowski:2007ft}. 

Initial-state fluctuations have a number of observable consequences in
nucleus-nucleus collisions: 
their relevance has first been pointed out in the study of the elliptic
flow, which is significantly enhanced by 
fluctuations~\cite{Miller:2003kd,Alver:2006wh,Andrade:2006yh}. 
They also generate odd harmonic components of anisotropic
flow~\cite{Luzum:2011mm}, such as the triangular flow~\cite{Alver:2010gr} and 
the directed flow at
midrapidity~\cite{Teaney:2010vd,Luzum:2010fb,Retinskaya:2012ky}. 
These new flow phenomena have been analyzed 
at Relativistic Heavy-Ion Collider (RHIC)~\cite{Adare:2011tg,Adamczyk:2013waa} and
at the Large Hadron Collider (LHC)~\cite{ALICE:2011ab,Chatrchyan:2012wg,ATLAS:2012at}. 

The recognition that initial-state fluctuations act as a seed for
anisotropic flow has triggered searches of collective flow in much
smaller systems, such as in the proton-nucleus
collisions~\cite{Bozek:2011if,Bzdak:2013zma},
where collective behavior could explain the
observed 
correlations~\cite{Abelev:2012ola,Adare:2013piz,Aad:2013fja,Chatrchyan:2013nka}.  
Anisotropic flow is also a
candidate~\cite{Werner:2010ss,Bozek:2010pb,Deng:2011at}
among others~\cite{Diehl:2011tt,Dusling:2013oia} for explaining the 
``ridge'' in the high-multiplicity proton-proton collisions at the
LHC~\cite{Khachatryan:2010gv,Li:2012hc}. 

The relevance of event-by-event
fluctuations~\cite{Heiselberg:2000fk,Jeon:2003gk} extends beyond the
realm of anisotropic flow. 
In particular, comprehensive studies have been devoted to 
fluctuations of the average transverse momentum $\langle p_T \rangle$,
both from the 
theoretical~\cite{Gazdzicki:1992ri,Broniowski:2005ae,Gavin:2006xd,Sharma:2008qr} 
and experimental side 
\cite{Adler:2003xq,Anticic:2008aa,Adamova:2008sx,Agakishiev:2011fs},
as these fluctuations may reflect critical
phenomena expected at the phase transition. 

Much progress has been made in understanding the response to initial
fluctuations~\cite{Holopainen:2010gz,Teaney:2010vd,Qiu:2011iv,Gardim:2011xv,Teaney:2012ke,Niemi:2012aj}:
typically, elliptic flow is proportional to the initial
eccentricity~\cite{Alver:2006wh}; 
triangular flow is proportional to the initial
triangularity~\cite{Alver:2010gr};
finally, $\langle p_t\rangle$ fluctuations arise as a natural
consequence of 
initial-state (size) fluctuations~\cite{Broniowski:2009fm}.
Eventually, the study of fluctuations 
essentially boils down to the study of the initial fluctuations. 

There have been many dedicated studies of the initial fluctuations using 
the  Glauber
approach~\cite{Broniowski:2007ft,Alvioli:2011sk,Staig:2010pn,Nagle:2010zk,Lacey:2010av,Qin:2011uw,Jia:2012ma},
as well as other approaches inspired by the saturation
physics~\cite{Schenke:2012hg,Dumitru:2012yr}. 
Partonic correlations, which are neglected here, have also been studied in
proton-proton and proton-nucleus collisions~\cite{Jung:2009eq,Calucci:2013pza}.
In other studies of fluctuations, the Gaussian Color Glass Condensate model
was considered in~\cite{Muller:2011bb} and the Fourier-Bessel fluctuating mode
decomposition was applied in~\cite{Floerchinger:2013rya,Floerchinger:2013vua}.

In the Glauber model, each wounded
nucleon~\cite{Bialas:1976ed,Bialas:2008zza}  is treated as a localized 
source. The resulting correlations are of two types: 
\begin{itemize}
 \item the correlations already present in the colliding nuclei, due in
   particular to the short-range nucleon-nucleon repulsion
(Sec.~\ref{sec:NNcorrelations}); 
 \item the correlations generated by the collision mechanism itself:
   a projectile nucleon can collide with a target nucleon only if they
   are close by in the transverse coordinate space, therefore wounded 
   nucleons go in pairs. We refer to this effect as to the
   ``twin'' correlations. 
\end{itemize}
While the first effect has already been thoroughly
studied~\cite{Broniowski:2010jd,Alvioli:2011sk}, 
there are fewer studies of twin correlations~\cite{Alver:2008zza}.

Most effects of initial-state fluctuations are encoded in the two-body
correlation of the initial  (energy)\footnote{Throughout this  paper we refer to the {\em energy density} simply as the {\em density}. We shall also assume that $\rho$ is divided by a constant energy factor, so as to give it the dimension of a number density.}
density $S(\x,\y)$~\cite{Baym:1995cz,Heiselberg:2000fk,Muller:2011bb}. 
This quantity is defined in Sec.~\ref{sec:def}, 
where we explain how fluctuations of observables can be expressed in
terms of $S(\x,\y)$. 
In Sec.~\ref{sec:NNcorrelations}, we introduce a simple
parametrization of nuclear correlations and show that their effect 
is reduced by the projection onto the transverse plane, which results
from the collision geometry at ultrarelativistic energies. 
Proton-nucleus collisions are studied in Sec.~\ref{sec:anmod} and 
nucleus-nucleus collisions in Sec.~\ref{sec:PbPb}. 
We carry out numerical simulations 
with GLISSANDO~\cite{Broniowski:2007nz,Rybczynski:2013yba}, 
a flexible code for Monte Carlo Glauber~\cite{Alver:2008aq}
calculations. 
Since we wish to focus on situations where the initial fluctuations
are most relevant, we only study central collisions, with exactly zero
impact parameter ($b=0$). 
Along with the first systematic investigation of twin correlations
(Sec.~\ref{sec:PbPb}), 
we  present semi-analytic estimates of effects 
of nuclear correlations, which have been investigated 
numerically in greater detail by other
groups~\cite{Broniowski:2010jd,Alvioli:2011sk}.  

\section{From correlations to fluctuations\label{sec:def}}

In this section, we first define the simple Glauber model which is used throughout 
this paper. 
We introduce the density-density correlation function $S(\x,\y)$ and 
show it decomposition. 
We then explain how fluctuations of 
observables can be expressed in terms of $S(\x,\y)$. 

\subsection{Glauber models \label{sec:glauber}}

In Glauber models, a proton-nucleus (p-A) or a nucleus-nucleus (A-A)
collision is viewed as a superposition of elementary processes, 
each of which deposits entropy and energy
locally~\cite{Broniowski:2007ft} in ``sources''. 
The simplest implementation is the wounded nucleon
model of A-A collisions~\cite{Bialas:1976ed,Bialas:2008zza}: 
nucleons from the colliding nuclei wound whenever their transverse distance is
sufficiently small, 
and point-like sources are created at the centers of wounded nucleons. 
For p-A collisions, we use an alternative
prescription~\cite{Bzdak:2013zma}, 
where the point-like sources are
located in the center-of-mass of the incident proton and the 
wounded nucleons from the target nucleus. 
Note that we choose different prescriptions for
the nucleus-nucleus and proton-nucleus collisions.
For the nucleus-nucleus collisions, this is the standard prescription. 
For the proton-nucleus collisions, our choice is dictated by simplicity, 
as explained later in this paper. 

In a given event, the density of sources in the transverse plane is 
\begin{eqnarray}
\label{eq:deltasources}
\rho(\x)=\sum_{i=1}^n \delta(\x-\x_i), 
\end{eqnarray}
where $n$ is the number of sources and $\x_i$ denote their transverse
positions. 
The integrated density is the number of sources: 
\begin{eqnarray}
\label{eq:normrho}
 \int \rmd^2\x\, \rho(\x)=n. 
\end{eqnarray}
The number $n$ fluctuates from event to event, and so do the positions $\x_i$. 
Since $n$ is related to the
multiplicity of the event, we shall refer to it simply as the
multiplicity.

Eq.~(\ref{eq:deltasources}) defines the simplest form of the Glauber
model, which is used throughout this paper. It can easily be improved
by taking into account  the fact that sources may not all
be equivalent~\cite{Broniowski:2007nz} (Appendix~\ref{app:sup}), or
by incorporating the finite transverse size of the sources 
(Appendix~\ref{sec:smear}).
Further ramifications
take into account the number of binary 
nucleon-nucleon collisions~\cite{Kharzeev:2000ph,Back:2001xy,Broniowski:2007nz},
or include a fluctuating nucleon-nucleon cross
section~\cite{Frankfurt:2008vi,Alvioli:2013vk,Rybczynski:2013mla,Alvioli:2014sba,Coleman-Smith:2013rla}.

Note that Monte Carlo Glauber calculations frequently involve 
recentering corrections: typically, one imposes $\sum_{i=1}^n\x_i=0$
in Eq.~(\ref{eq:deltasources}). The correlation induced by this recentering 
correction is discussed in Appendix~\ref{sec:recenter}. 

\subsection{Density-density correlations\label{sec:densitydensity}}

The central object of our study is the density-density correlation
function, defined as
\begin{eqnarray}
S(\x,\y)&\equiv&\langle \rho(\x) \rho(\y) \rangle - \langle \rho(\x) \rangle \langle \rho(\y) \rangle, \nonumber\\
&=&  \langle \delta\rho(\x) \,\delta\rho(\y)
\rangle \label{eq:defcor}, 
\end{eqnarray}
where $\langle\cdots \rangle$ denotes the average over a large number
of events, 
and $\delta\rho(\x)\equiv\rho(\x)-\langle\rho(\x)\rangle$ is the
fluctuation at a given point around the average 
density $\langle\rho(\x)\rangle$. 

From Eq.~(\ref{eq:normrho}), one derives the normalization
\begin{eqnarray}
\int \rmd^2\x \rmd^2\y \,S(\x,\y) = {\rm Var}(n).\label{eq:Snorm}
\end{eqnarray}
Using Eq.~(\ref{eq:deltasources}), one can put $S(\x,\y)$ in the form 
(see Appendix~\ref{sec:cps}) 
\begin{eqnarray}
 S(\x,\y) = \langle \rho(\x) \rangle \delta(\x-\y) + \langle \rho(\x) \rangle \langle \rho(\y) \rangle [g(\x,\y)-1]. \nonumber \\ \label{eq:ssxy}
\end{eqnarray}
The first term in the right-hand side is the so-called autocorrelation, which is
proportional to $\delta(\x-\y)$ for point-like sources: it is the
contribution of density fluctuations.  
The remaining terms involve the standard {\em pair distribution
  function} $g(\x,\y)$ (cf. Eq.~(\ref{eq:gdef})), which contains the
information on correlations between sources. 
The decomposition (\ref{eq:ssxy}) can be generalized to the case of 
sources of fluctuating strength (Appendix~\ref{app:sup}) and sources
of finite size (Appendix~\ref{sec:smear}). 

Many analyses in heavy-ion collisions are done at a fixed centrality,
where the centrality is typically determined according to the
multiplicity. 
Within our simple Glauber model, this amounts to
fixing the number of sources, therefore our simulations in this
paper are always carried out for fixed $n$.
{Note that we fix both 
the impact parameter $b=0$ and the number of sources $n$ throughout the paper.
This is done in practice by randomly generating events with $b=0$, and then accepting events with a 
given value of $n$. The purpose to fix $b$ (which can be done in simulated
events) is to eliminate the extra fluctuations originating from the impact
parameter, which obscure the mechanisms we wish to point out.} 
If the sources are uncorrelated 
(which is the case considered in~\cite{Bhalerao:2006tp,Bhalerao:2011bp}),
the pair distribution function reduces to 
\begin{eqnarray}\label{eq:gnocorr}
g(\x,\y) =1-\frac{1}{n}. 
\end{eqnarray}
By inserting this expression into Eq.~(\ref{eq:ssxy}) one obtains 
\begin{equation}
\label{eq:ssxynocorr}
 S(\x,\y) = \langle \rho(\x) \rangle \delta(\x-\y) -\frac{1}{n} \langle \rho(\x) \rangle \langle \rho(\y) \rangle.
\end{equation}
The density-density correlation thus reduces to the autocorrelation
term, minus a compensating term which ensures that it integrates to 0, as
required by Eq.~(\ref{eq:Snorm}). 

\subsection{Fluctuations of observables \label{sec:res}}

We now explain how fluctuations of observables relate to the
density-density correlation function $S(\x,\y)$. 
Observables are determined, to a good approximation, by simple
properties of the initial density profile ---which we also refer to as
``observables'' by a slight abuse of terminology--- 
such as the mean squared radius
\begin{equation}
\label{eq:defR}
r_{\rm rms}^2\equiv\frac{\int \rmd^2 \x\, r^2 \rho(\x)}{\int \rmd^2 \x\,\rho(\x)},
\end{equation}
and the initial anisotropy $\varepsilon_n$ in harmonic
$n$~\cite{Teaney:2010vd}, with $n\ge 2$:
\begin{equation}
\label{eq:defepsilon}
\varepsilon_n \equiv 
\frac{\int \rmd^2 \x\, r^ne^{in\phi} \rho(\x)}{\int \rmd^2
  \x\,r^n\rho(\x)},
\end{equation}
where $(r,\phi)$ are polar coordinates for the transverse position
$\x$. 
Note that $\varepsilon_n$ thus defined is a complex number, whose modulus is the usual 
anisotropy, and whose phase yields the participant plane angle in harmonic $n$. 

In this paper, we use slightly different definitions of observables, in the sense
that we replace $\rho(\x)$ by $\langle\rho(\x)\rangle$ in the
denominators of Eqs.~(\ref{eq:defR}) and (\ref{eq:defepsilon}). 
The denominator of Eq.~(\ref{eq:defR}) is the number of sources $n$, 
which is kept fixed in all our simulations, and therefore coincides with its mean $\langle n\rangle$. 
For Eq.~(\ref{eq:defepsilon}), the argument is different: 
The numerator vanishes by symmetry  if one replaces $\rho(\x)$ by
$\langle\rho(\x)\rangle$ for central collisions. 
Therefore, if one replaces $\rho(\x)$ by
$\langle\rho(\x)\rangle$ in the denominator, $\varepsilon_n$ is unchanged to
leading order in the fluctuations $\delta\rho(\x)$.
We have checked numerically that the difference is irrelevant in
practice. 
The practical purpose of averaging separately the numerators and
denominators in Eqs.~(\ref{eq:defR},\ref{eq:defepsilon}) is that these 
expressions are linear in $\rho(x)$. Then parts of the analysis can be carried
our analytically. 

Strictly speaking, Eqs~(\ref{eq:defR}) and (\ref{eq:defepsilon}) hold
in a centered coordinate system, such that 
$\int \rmd^2 \x\, \x\rho(\x)=0$ for every event. 
In this section, we neglect this recentering correction and only
center the average distribution: $\int \rmd^2 \x\,
\x\langle\rho(\x)\rangle=0$.  
The recentering correction is a higher-order correction to the size,
and also to anisotropies for central
collisions~\cite{Bhalerao:2011bp}, except for the dipole asymmetry
$\varepsilon_1$ which is not studied here~\cite{Teaney:2010vd}.  

With these approximations, observables are determined by
simple integrals of the density profile of the type
\begin{equation}
{\cal O}\equiv\int \rmd^2 \x\, \Omega(\x) \rho(\x), 
\end{equation}
with $\Omega(\x)=r^2,r^ne^{in\phi}$. 

The mean value of  ${\cal O}$ is obtained by replacing 
$\rho(\x)$ with $\langle\rho(\x)\rangle$ in this equation. One thus 
obtains
\begin{eqnarray}
\langle r_{\rm rms}^2\rangle &=& \frac{1}{n}\int \rmd^2 \x\, r^2 \langle\rho(\x)\rangle\cr
\langle \varepsilon_n\rangle&=& 0.
 \label{eq:meanvalues}
\end{eqnarray}
The mean anisotropy is 0 for all $n$ because we are considering central collisions.

Similarly, the variance of ${\cal O}$ is readily expressed as a function of
the density-density correlation (\ref{eq:defcor}):
\begin{equation}
\label{eq:variances}
{\rm Var}({\cal O})
=\frac{1}{n^2}\int \rmd^2 \x\, \rmd^2\y\, \Omega(\x)\Omega^*(\y) S(\x,\y).
\end{equation}

The uncorrelated case, defined by Eq.~(\ref{eq:gnocorr}), will often
serve as  a benchmark. 
Inserting Eq.~(\ref{eq:ssxynocorr}) into Eq.~(\ref{eq:variances}), one
obtains  
\begin{eqnarray}
{\rm Var}(r_{\rm rms}^2)^{\rm no~corr.} &=& \frac{1}{n}\left(\langle r^{4}\rangle- \langle r^2 \rangle^2\right),\cr
{\rm Var}(\varepsilon_n)^{\rm no~corr.} &=& 
\frac{\langle r^{2n} \rangle}{ n\langle r^n \rangle^2}, \; \label{eq:eincnc}
\end{eqnarray}
where 
\begin{equation}
\langle r^n \rangle\equiv \frac{1}{n}\int \rmd^2
  \x\,r^n\langle\rho(\x)\rangle
\end{equation}
are the moments of the $r$ distribution with the average density profile
$\langle\rho(\x)\rangle$
(note that $\langle r^2\rangle=\langle r_{\rm rms}^2\rangle$).
These results (\ref{eq:eincnc}) are already known for uncorrelated
sources~\cite{Bhalerao:2011bp}. 
Note that the last term in Eq.~(\ref{eq:ssxynocorr}), which is
disconnected, does not contribute to the $\varepsilon_n$ fluctuations,
which are solely due to the autocorrelation term.

One of our goals in this paper is to carefully evaluate the effect of
correlations on fluctuation measures. 
To this end, we introduce the ratio
\begin{equation}
R({\cal O})=\frac{{\rm Var}({\cal O})}{{\rm Var}({\cal O})^{\rm no~corr}} \label{eq:ratioR}
\end{equation}
which we will evaluate for the proton-nucleus in Sec.~\ref{sec:anmod} and for
the nucleus-nucleus collisions in Sec.~\ref{sec:PbPb}. 

\section{Nuclear correlations \label{sec:NNcorrelations}}

It is well known that the nucleons  inside nuclei are strongly
correlated, making nuclear matter behave more as a Fermi liquid than
a Fermi gas. 
In this section, we recall their effects on the Glauber
calculations. These effects have been studied in detail in 
Refs.~\cite{Broniowski:2010jd,Alvioli:2011sk}.   

Sources in the Glauber model are associated with wounded nucleons,
projected onto the transverse plane (Eq.~(\ref{eq:deltasources})).
Correlations between sources stem from correlations between wounded
nucleons, hence they are affected by nuclear correlations. 
We are going to show that these effects are 
significantly reduced by the projection onto the transverse plane.

\begin{figure}[tb]
\begin{center}
\includegraphics[angle=0,width=0.45 \textwidth]{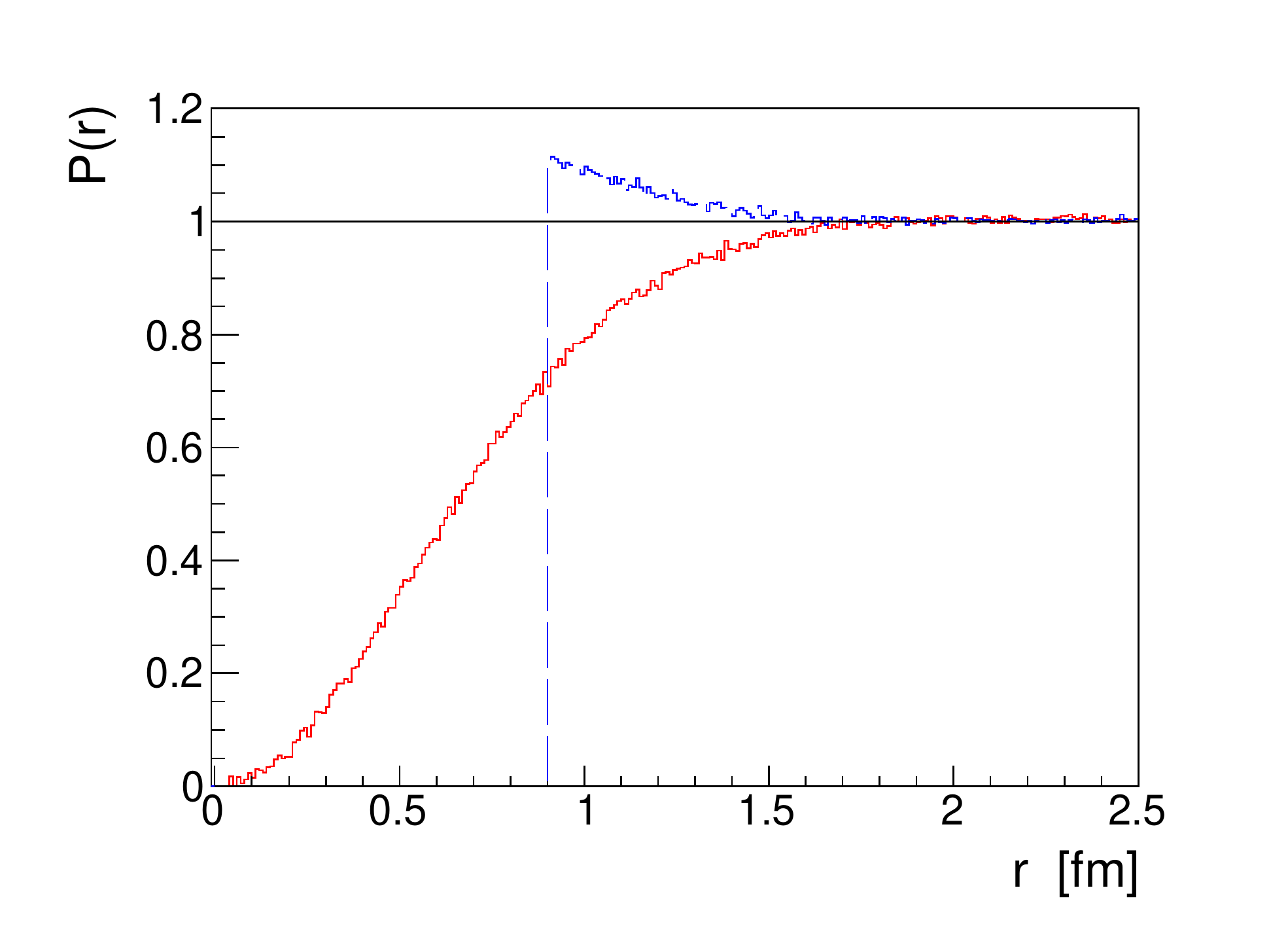} 
\end{center}
\vspace{-7mm}
\caption{(Color online) The pair distribution function of Eq.~(\ref{eq:defPr})
in the relative distance, $P(r)$, 
for the $^{208}$Pb nucleus for the hard-sphere expulsion (dashed line) and Gaussian 
correlation (solid line), obtained from a Monte Carlo simulation with
GLISSANDO~\cite{Broniowski:2007nz} 
using the expulsion distance. \label{fig:ar}}
\end{figure}    

The strong repulsive character of the nucleon-nucleon (NN) interaction
at short distances, together with the Pauli exclusion principle, generate
short range correlations. 
In this section, we model these with a hard sphere repulsion  or a 
smoother Gaussian repulsion which mimics the
distributions of Ref.~\cite{Alvioli:2009ab}. 

A natural measure of the correlation is the ratio between the
normalized two-body probability distribution of nucleons and the
product of the one-body distributions 
$f^{(2)}(\x_1,\x_2)/f^{(1)}(\x_1)f^{(1)}(\x_2)$ (see
Appendix~\ref{sec:cps}). 
In order to see how it varies as a function of the relative distance 
$r=|\x_1-\x_2|$, we integrate both the numerator and denominator over
the mean point:
\begin{equation}
\label{eq:defPr}
P(r)\equiv\frac
{\int\rmd^3\R\, f^{(2)}(\R+\frac{\r}{2},\R-\frac{\r}{2})}
{\int\rmd^3\R\, f^{(1)}(\R+\frac{\r}{2})f^{(1)}(\R-\frac{\r}{2})}
\end{equation}
The half-integrated pair correlation function $P(r)$ thus defined is
plotted in Fig.~\ref{fig:ar} for the hard sphere and the 
Gaussian repulsion. Note the strong repulsion dip near the origin,
compensated by a slight overshoot above unity at large distances. 

\begin{figure}[tb]
\begin{center}
\includegraphics[angle=0,width=0.45 \textwidth]{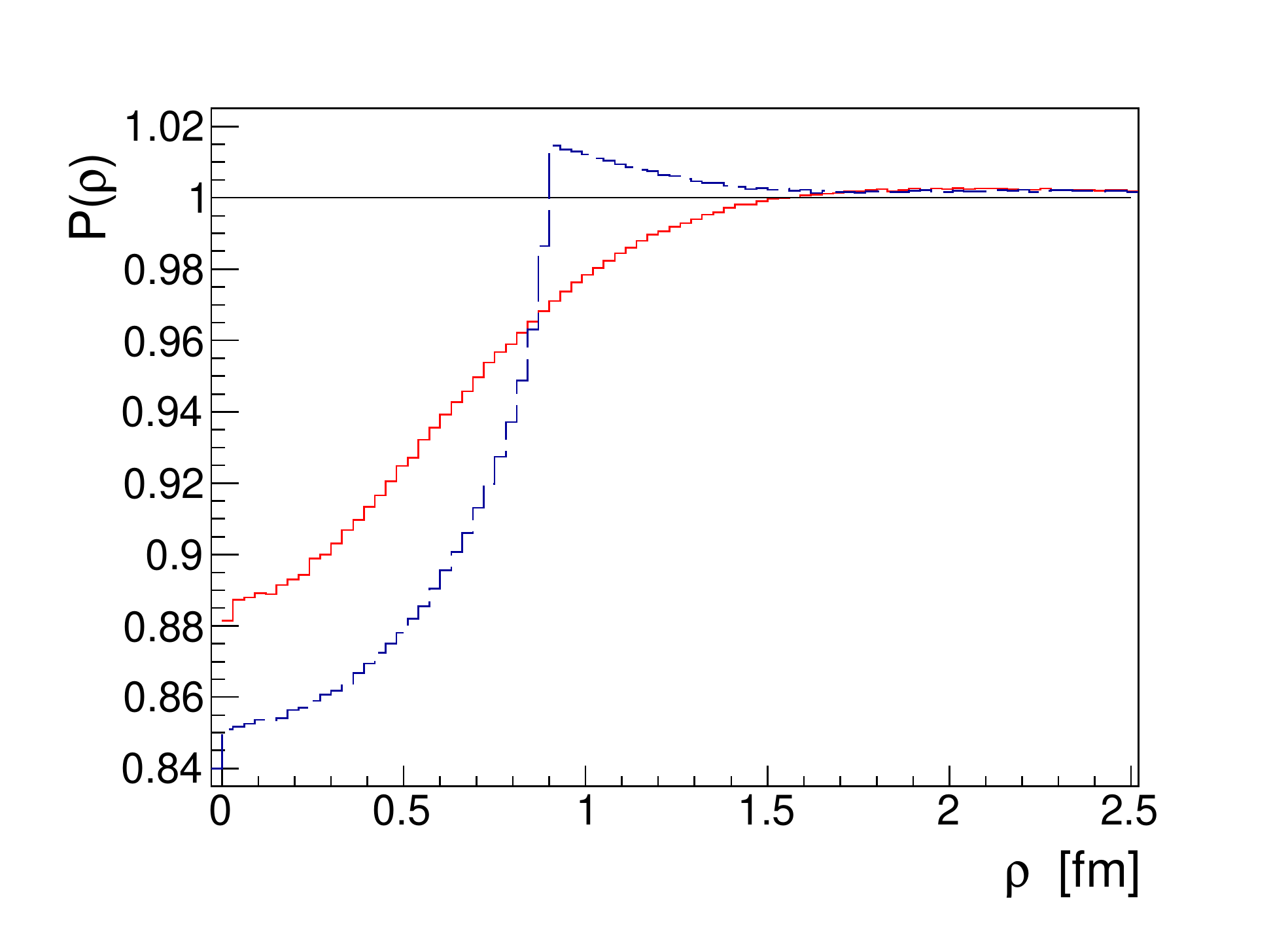} 
\end{center}
\vspace{-7mm}
\caption{(Color online) Same as Fig.~\ref{fig:ar} but for the distribution in
the transverse distance, $P(\rho)$ ($\rho=s$).
\label{fig:arho}}
\end{figure} 
The relevant quantity for the present study, however,  is 
the pair correlation function in the relative {\it transverse\/} distance,
obtained by integrating both the numerator and the denominator of
Eq.~(\ref{eq:defPr}) along the collision axis $z$ (in the same way 
as the nuclear thickness function~\cite{Miller:2007ri}). 
The resulting pair correlation function $P(s)$ is shown in Fig.~\ref{fig:arho}.
We note that the projection leading from $P(r)$ to $P(s)$ greatly reduces the repulsion dip, from $1$ to $\lesssim 0.15$. This is easy to 
understand, since the two nucleons may lie close to each other in the
transverse plane, i.e., be at small $s$, while being sufficiently
separated in $z$, and hence outside the volume over which they feel
the short range correlation. This property will allow us to treat these
correlations as a small perturbation. 

We thus write $P(s)=1-d(s)$, where $d(s)\ll 1$. In the following
sections, we use the Gaussian parametrization of the nuclear
repulsion, which is more realistic than the hard-sphere
repulsion~\cite{Rybczynski:2011wv}. The resulting $d(s)$ is itself
Gaussian to a good approximation: 
\begin{eqnarray}
&& d(s)= B  \exp \left (-\frac{s^2}{2\sigma_d^2} \right ), \label{eq:pard} \\
&& B=0.11, \;\; \sigma_d=0.56~{\rm fm},\nonumber 
\end{eqnarray}
where the numerical values have been fitted to the results shown in 
Fig.~\ref{fig:arho}. 
When comparing the cases of nuclear distributions with and without 
the NN correlations, we make sure that the single-particle
distributions $f^{(1)}(\x)$ are identical. 
The way to accomplish this in Monte Carlo studies with NN expulsion distance is
explained in
Ref.~\cite{Broniowski:2007nz}.  

\section{Proton-nucleus collisions \label{sec:anmod}}

In this section, we study the one-body density and the density-density
correlation in the proton-nucleus collisions analytically and
numerically. We then evaluate fluctuation observables. 

We consider central proton-nucleus collisions ($b=0$). We denote with
$\theta(s)$ the {\em wounding profile}, i.e., 
the probability that the proton  
interacts inelastically with a nucleon sitting at a transverse
distance $s$ away from it. 
We use the following Gaussian parametrization~\cite{Alvioli:2009ab}
\begin{eqnarray}
&& \theta(s)=A \exp \left (-\frac{s^2}{2\sigma_w^2} \right ), \label{eq:parth} \\
&& A=0.92, \;\; \sigma_w=1.08~{\rm fm}. \nonumber 
\end{eqnarray}
The normalization reproduces the total inelastic NN cross section,
i.e., $\int \rmd s\, \theta(s)=2\pi A\sigma_w^2= \sigma_{NN}^{\rm inel}$.  
The choice of parameters in Eq.~(\ref{eq:parth}) is such that
$\sigma_{NN}^{\rm inel}\simeq 68$~mb, corresponding to the LHC energy
$\sqrt{s_{NN}}=5.02$~TeV. 
Nucleons hit by the proton as it crosses the nucleus are referred to
as participants or wounded nucleons.

In the model considered here~\cite{Bzdak:2013zma}, a source is produced at
 mid-distance between the proton and each participant nucleon. 
Since the proton is by assumption located at the origin, the sources are 
placed at the positions $\z_i\equiv \s_i/2$, where $\s_i$ denotes the
transverse location of the $i^{\rm th}$ participant. 
Thus there is a one-to-one correspondence between sources and hit nucleons in 
the target, and all the sources play symmetric roles. 
Another option is to assume that the sources are created on top of each
participant. The resulting differences are discussed briefly at the end of
Sec.~\ref{sec:analytic}. 

\subsection{Analytic model \label{sec:analytic}}

As seen above,  
the distribution of sources closely follows that of the 
participants within the target nucleus;  in particular, this
distribution directly reflects
nuclear correlations that are inherited from those among
the nucleons of the target nucleus. 
As explained in Appendix~\ref{sec:pAanalytic}, for weak 
correlations, the two-body
distribution of the sources is approximately given by (with $\s_i=2\z_i$)  
\begin{equation}\label{eq:f2gl}
f^{(2)}(\z_1,\z_2)\simeq 
c_2 \,{\theta(\s_1) \theta(\s_2) (1-d(|\s_1-\s_2|))},
\end{equation}
where $c_2$ is a normalization constant. 
It is thus entirely  determined by the wounding profile $\theta$ and the
correlation factor $1-d$.

The one-body distribution $f^{(1)}(\z_1)$ 
is obtained by integrating Eq.~(\ref{eq:f2gl}) over $\z_2$ 
(see Eq.~(\ref{eq:f1f2})).  
For uncorrelated nucleons ($d(|\s_1-\s_2|)=0$), it reduces to 
\begin{equation}
\label{eq:f1nocorr}
f^{(1)}(\z_1)=\frac{1}{n}\langle\rho(\z_1)\rangle\propto 
\theta(\s_1). 
\end{equation}
The one-body distribution thus reproduces the wounding profile
(\ref{eq:parth}). 
When the nucleon-nucleon correlations are taken into account,
the one-body distribution of the
{\it sources\/} changes slightly, even though the one-body distribution of
{\it nucleons\/} is strictly unchanged (see the statement at the very end of
Sec.~\ref{sec:NNcorrelations}).
However, this is a very tiny change in practice, such that
Eq.~(\ref{eq:f1nocorr}) holds to a very good approximation with
realistic correlations.

\begin{figure}[tb]
\begin{center}
\includegraphics[angle=0,width=0.45 \textwidth]{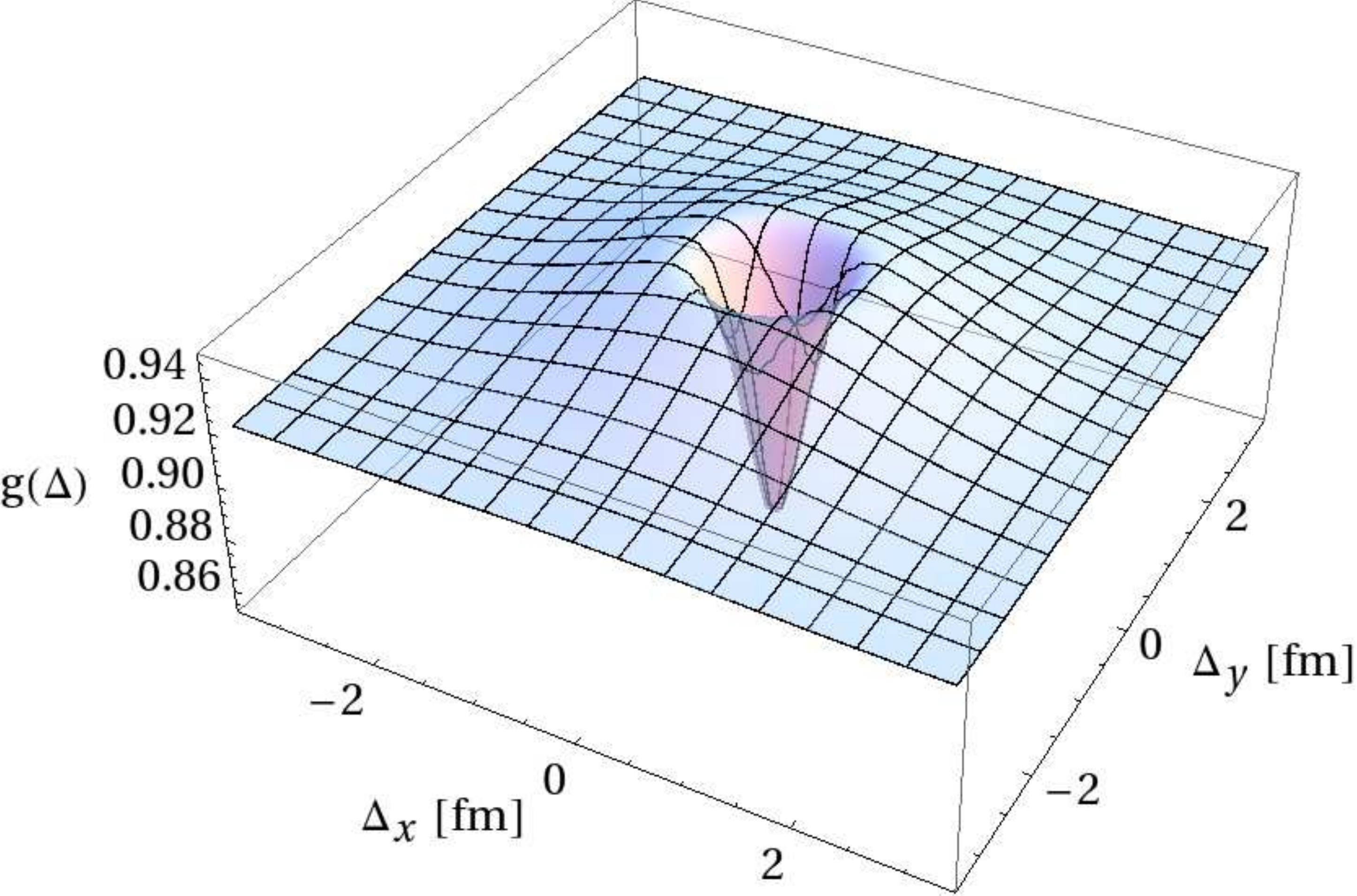} 
\end{center}
\vspace{-7mm}
\caption{(Color online) The
half-integrated pair distribution function 
$g(\Delta_x,\Delta_y)$ for the fireball created in  the p+Pb collisions at the
  impact parameter $b=0$ and the 
number of participants $n=15$, 
$\Delta$ is the relative transverse distance between the two sources.
\label{fig:fanalytg}}
\end{figure}   

 \begin{figure*}[tb]
\begin{center}
\includegraphics[angle=0,width=0.85 \textwidth]{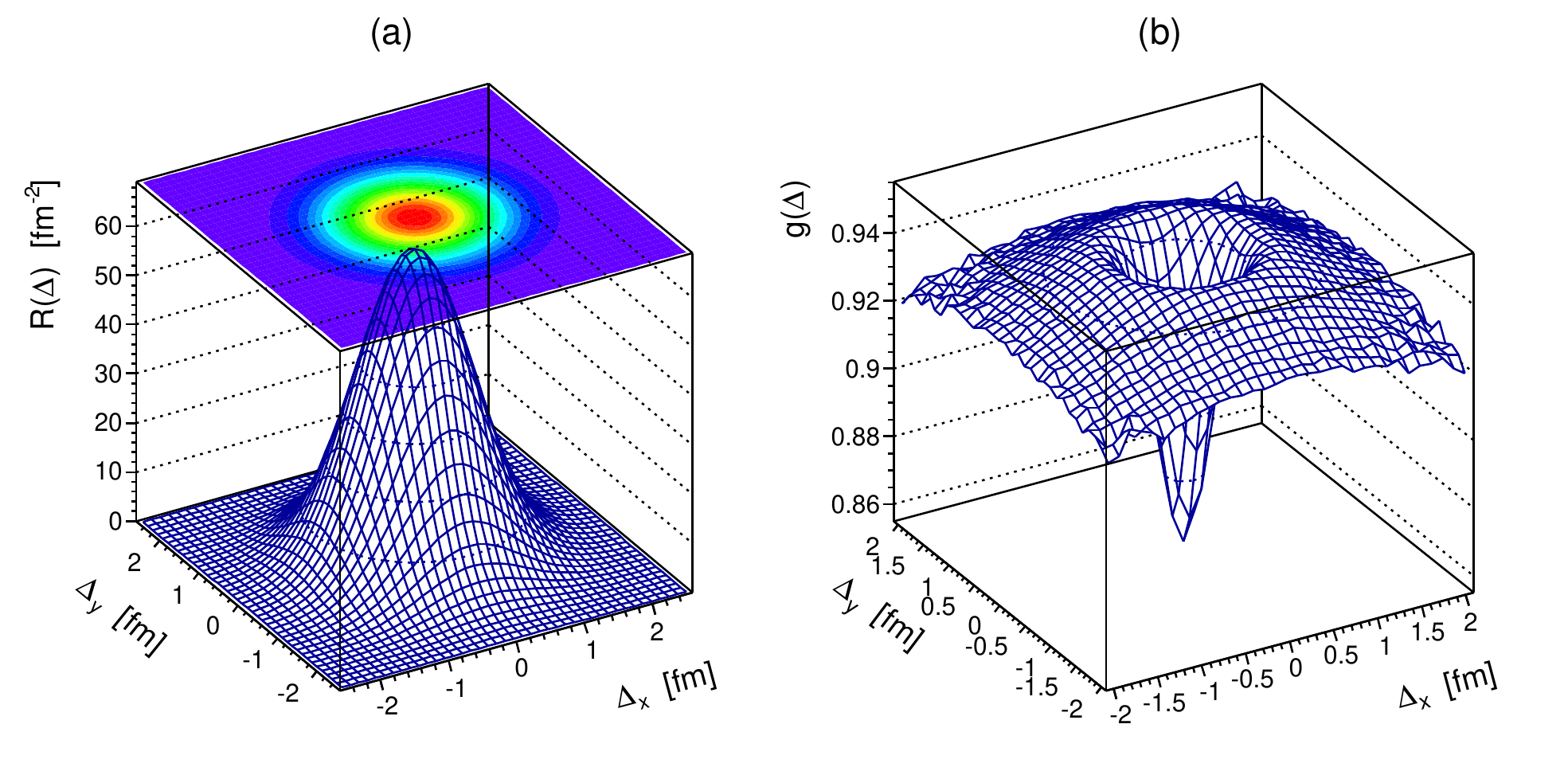} 
\end{center}
\vspace{-7mm}
\caption{(Color online) Functions $R(\Delta)$ (a) and $g(\Delta)$ (b) defined in Eq.~(\ref{eq:rgr}) 
for p+Pb collisions at the impact parameter $b=0$ and the number of participants
$n=15$, obtained with GLISSANDO.
\label{fig:f1}
}
\end{figure*}   

In order to study how the density-density correlation depends on the
relative distance, we introduce the {\em half-integrated} density-density 
correlation function, much in the same way as in Eq.~(\ref{eq:defPr}):
\begin{eqnarray}
S(\D)=\int \rmd\r\, S( \r+\frac{\D}{2}, \r-\frac{\D}{2} ) . \label{eq:rhoD}  
\end{eqnarray}
The decomposition (\ref{eq:ssxy}) can be easily transposed to the
half-integrated correlation: 
\begin{equation}
S(\D) =   \langle n\rangle \delta(\D ) + R(\D ) [g(\D )-1],
\end{equation}
where
\begin{eqnarray}\label{eq:rgr} 
R(\D ) &\equiv& \int \rmd\r\, \left\langle \rho( \r+\frac{\D}{2} )
  \right\rangle \left\langle \rho\left( \r-\frac{\D}{2}\right)
  \right\rangle, \cr
g(\D )&\equiv& \frac{1}{R(\D)} \int \rmd\r \, \left\langle n(n-1)f^{(2)}(\r+\frac{\D}{2},\r-\frac{\D}{2}) \right\rangle\nonumber\\ . 
\end{eqnarray}

With the approximation of Eq.~(\ref{eq:f2gl}) for the two-body distribution, 
all integrals can be evaluated analytically. 
The pair distribution function $g(\D)$ is plotted in
Fig.~\ref{fig:fanalytg} (right).  
The central dip reflects the short-range nucleon-nucleon repulsion.
As a crude approximation, by using Eq.~(\ref{eq:f1nocorr}), 
one may replace  $\theta(\s_i)$ by $f^1(\z_i)$ in Eq.~(\ref{eq:f2gl}), and write 
$f^{(2)}(\z_1,\z_2)\simeq f^{(1)}(\z_1)f^{(1)}(\z_2)(1-d(|\s_1-\s_2|))$. 
One thus obtains
\begin{equation}
\label{eq:gnonorm}
g(\D)\simeq \left(1-\frac{1}{n}\right)(1-d(2\Delta)), 
\end{equation}
where we have used the fact that the source is half-way between the
nucleons, $\s_i=2\z_i$. 
Eq.~(\ref{eq:gnonorm}) is a good approximation to the exact result in
Fig.~\ref{fig:fanalytg}. 
A more careful calculation, enforcing the proper normalization, 
shows that this approximation only holds when the range of
correlations is much smaller than the wounding profile,
$\sigma_d\ll\sigma_w$, a condition which is only marginally satisfied
with the chosen numerical values.

The simplicity of the results in 
Eqs.~(\ref{eq:f1nocorr}) and (\ref{eq:gnonorm}) 
is a consequence of the prescription
to locate sources half-way between the proton and the wounded
nucleons. 
If one chooses instead the ``standard'' prescription where sources are
located around each participant,
the proton also acts as a source in addition to the hit nucleons:
this results in cumbersome
expressions for the one-body and two-body densities, but no new
insight. 

\subsection{Monte Carlo simulations \label{sec:pPbsim}}

We now show that our simple analytic results, based on the approximation 
of weak correlations, are fully supported by numerical simulations made with GLISSANDO~\cite{Broniowski:2007nz}  
at zero impact parameter, $b=0$. 
The condition $b=0$ is not realistic, in the sense that it cannot be
implemented experimentally. On the other hand, it allows us 
to clearly isolate correlations which do not originate in impact
parameter fluctuations.  
For sake of simplicity, we fix the number of participants to $n=15$, 
corresponding to a typical value at the LHC energy. 
In the presented 
simulations we are using the correlated nuclear distributions of
Ref.~\cite{Alvioli:2009ab} 
and the Gaussian wounding
profile~\cite{Rybczynski:2011wv} of Eq.~(\ref{eq:parth}).
 
In Fig.~\ref{fig:f1} we show the two-dimensional plots of 
the simulated half-integrated correlation function $R(\Delta)$, and the pair correlation 
function $g(\Delta)$ of Eq.~(\ref{eq:rgr}). The function $R(\Delta)$ 
simply reflects the shape of the folding of the two single-particle distributions, while $g(\Delta)$ 
is remarkably close to the result of the analytic calculation presented in 
Fig.~\ref{fig:fanalytg}.

\subsection{Observables}

The approximations made in Sec.~\ref{sec:analytic} 
(specifically, Eqs.(\ref{eq:pard}), (\ref{eq:parth}) and
(\ref{eq:f2gl})) 
allow us to obtain
analytic expressions of observables introduced in Sec.~\ref{sec:res}. 
For a fixed number of participants $n$, the mean squared radius,
Eq.~(\ref{eq:defR}),
is 
\begin{eqnarray}
\label{eq:rmsPb}
\langle r_{\rm rms}^2\rangle&=&
\int \rmd^2 \x\, r^2 f^{(1)}(\x)\cr
&=&
\frac{\sigma _w^2}{2}\left(1+
   B\frac{ \sigma _d^2 \sigma _w^2}{ \left(\sigma _d^2+2 \sigma_w^2\right)^2}+{\cal O}\left(B^2\right)\right), \label{eq:R2}
\end{eqnarray}
where the one-body density $f^{(1)}(\x)$ has been obtained by integrating
Eq.~(\ref{eq:f2gl}). 
If nucleons are uncorrelated ($B=0$), the rms radius is proportional to the wounding radius 
$\sigma_w$. The proportionality constant would be typically a factor 2 larger if we had assumed that 
the sources were on top of the participants, instead of half-way between the proton and the participants
in the target nucleus. 

Equation~(\ref{eq:rmsPb}) shows that repulsive correlations ($B>0$) 
result in a very small increase of the fireball size.
The effect of correlations vanishes not
only  in the limit of a small correlation length, $\sigma_d\ll\sigma_w$, but 
also in the opposite limit where the correlation length is much larger than 
the fireball size, $\sigma_d\gg\sigma_w$.
This is a general result, as we shall see below for other observables. 
Numerically, the increase of  $\langle r_{\rm rms}^2\rangle$ due to 
repulsive correlations is a modest $0.6$\%. 

We now evaluate fluctuation observables, namely, the variance of the mean-square radius and the 
mean square eccentricity. In the uncorrelated case ($B=0$), 
Eq.~(\ref{eq:eincnc}) simplifies to 
\begin{eqnarray}
{\rm Var}(r_{\rm rms}^2)^{\rm no~corr.} &=& \frac{\langle r_{\rm rms}^2\rangle^2}{n},\cr
{\rm Var}(\varepsilon_2)^{\rm no~corr.} &=& 
\frac{2}{ n}, 
\end{eqnarray}
where we have used the fact that the one-body distribution, Eq.~(\ref{eq:f1nocorr}), is Gaussian.

We finally evaluate the changes in these observables due to correlations ($B>0$), 
which are conveniently expressed by the ratio of Eq.~(\ref{eq:ratioR}). 
Correlations modify all terms in the density-density correlation
function Eq.~(\ref{eq:ssxy}):
The autocorrelation term changes slightly due to the modification of the one-body 
distribution of sources. We neglect this contribution, which is typically less than 1\%, and 
evaluate the contribution of pair correlations. This contribution is enhanced by a
factor $n$ because the pair correlation $g(\x,\y)$ in Eq.~(\ref{eq:ssxy}) is multiplied by a the square of the average density, 
which scales as $n^2$, while the autocorrelation term only scales as $n$. 
After some algebra, one obtains 
\begin{eqnarray}
\label{eq:R4}
R(r_{\rm rms}^2) &=& 1 - 2nB
\frac{\sigma _d^2 \sigma _w^4}{\left(\sigma _d^2+2 \sigma _w^2\right)^3} +{\cal O}\left(B^2\right), \cr 
R(\varepsilon_2) &=& 1 - nB \frac{\sigma _d^2 \sigma_w^4 }
{\left(\sigma _d^2+2 \sigma _w^2\right)^3}+{\cal O}\left(B^2\right).
\end{eqnarray}
Numerically, correlations decrease $\langle|\varepsilon_2|^2\rangle$ by a modest  4\%. 
We have also evaluated numerically their effect of $\langle|\varepsilon_3|^2\rangle$ 
which is even smaller, at the level of 1\%.
Generally, effects of correlations are usually even smaller when
realistic NN correlations are implemented~\cite{Alvioli:2011sk},
rather than just the short-range part considered here.    
We conclude that for the proton-nucleus collisions  in the Glauber model,
correlations between sources are essentially negligible for the considered
observables.

\section{Nucleus-nucleus collisions \label{sec:PbPb}}

The Glauber~\cite{Czyz:1969jg,Bialas:1976ed,Bialas:2008zza}  description of the A+A collisions is inherently more complicated than 
that of the p+A collisions. We are no longer able  to determine analytically the
distribution of sources, and hence deduce their correlations with analytic
tools.  Nevertheless, one can get a simple understanding  of these correlations
in the two limiting cases of small and large inelastic cross sections. Further
understanding is gained through GLISSANDO simulations.

\subsection{Twin correlations \label{sec:twin}}

In the wounded-nucleon model, the density is created by participant nucleons, which come from one of the two colliding nuclei $A$ and $B$. 
By definition, to each participant in nucleus $A$ corresponds at least one
participant in nucleus $B$, at a distance of the order $\sqrt{\sigma_{NN}^{\rm
inel}}$ in the transverse plane. 
This condition creates nontrivial correlations between participants, which we dub ``twin'' 
correlations.
The effect of the twin correlations can be easily understood in the limits of very small or very large wounding cross section $\sigma_{NN}^{\rm inel}$.

For this discussion, it is useful to separate the contributions of the sources coming from nuclei $A$ and $B$, respectively, 
and write the total density of  sources
as $\rho(\x)=\rho_A(\x)+\rho_B(\x)$.
In the  limit of small $\sigma_{NN}^{\rm inel}$, each nucleon of $A$ sees at
most one nucleon of $B$, i.e., the wounded nucleons come in pairs, one from
nucleus $A$ and the other one from nucleus $B$. Furthermore, the smallness of
the cross section also implies that  both nucleons in the pair are close to
each other in the transverse plane. Thus, if there is a source from nucleus $A$
at point $\x$, there is also a source from nucleus $B$ at the almost same point
$\x$, 
therefore $\rho_A(\x)\simeq\rho_B(\x)$.
One can take into account the correlations among sources within
nucleus $A$ by defining a density-density correlation $S_{A}(\x,\y)$, 
and decomposing it according to Eq.~(\ref{eq:ssxy}):
\begin{eqnarray}
 S_{A}(\x,\y) &=& \langle \rho_A(\x) \rangle \delta(\x-\y) \cr
 &&+\langle \rho_A(\x) \rangle \langle \rho_A(\y) \rangle [g_A(\x,\y)-1]. \label{eq:ssxyA}
\end{eqnarray}
Since $\rho(\x)\simeq 2\rho_A(\x)$, the full density-density
correlation function is $S(\x,\y)\simeq 4S_A(\x,\y)$, and the previous equation gives
\begin{eqnarray}
 S(\x,\y) &=&2 \langle \rho(\x) \rangle \delta(\x-\y) \cr
 &&+\langle \rho(\x) \rangle \langle \rho(\y) \rangle [g_A(\x,\y)-1]. \label{eq:ssxytwin}
\end{eqnarray}
Comparing with the general decomposition (\ref{eq:ssxy}), one sees that the twin correlations double the autocorrelation term. 
Equivalently, they give a $\delta$ contribution to the pair distribution function:
\begin{equation}
g(\x,\y)=g_A(\x,\y)+\frac{1}{\langle\rho(\y)\rangle}\delta(\x-\y).  \label{eq:twin}
\end{equation}

\begin{figure}[tb]
\begin{center}
\includegraphics[angle=0,width=.5 \textwidth]{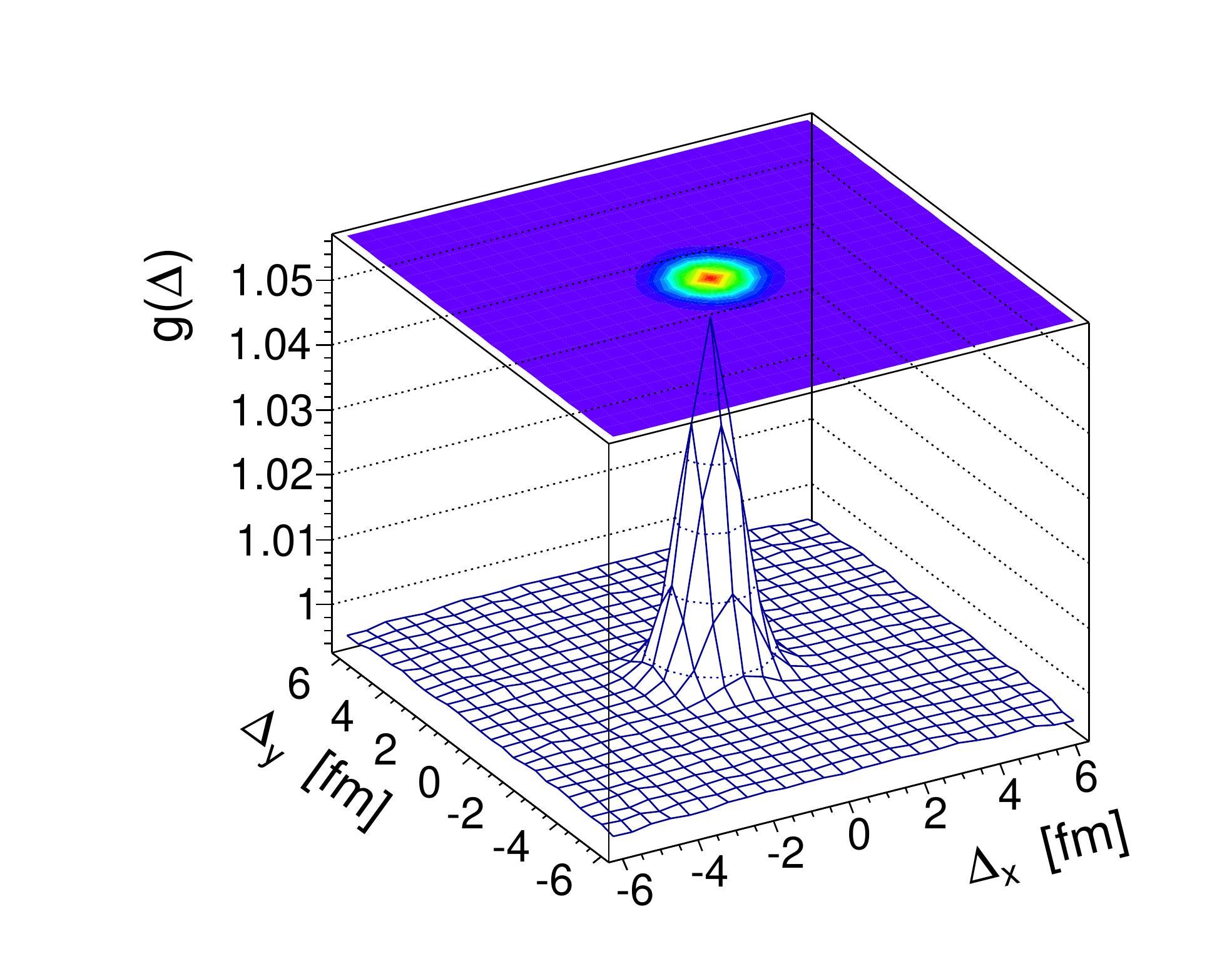} 
\end{center}
\vspace{-2mm}
\caption{(Color online) The half-integrated pair correlation function $g(\Delta)$ 
for Pb-Pb collisions at impact parameter $b=0$ 
with $\sigma_{NN}^{\rm inel}=20$~mb and $n=371$. 
No NN repulsion in the nuclear distributions is included. 
A sharp peak from the twin-production mechanism is clearly visible.
\label{fig:f20}
}
\end{figure}

Fig.~\ref{fig:f20} illustrates this result with a Monte Carlo Glauber simulation
carried out for central ($b=0$) Pb-Pb collisions with a value of the cross
section much lower than the actual value at the LHC, namely, $\sigma_{NN}^{\rm
inel}=20$~mb.
We use a Gaussian wounding profile. 
In order to eliminate effects of multiplicity fluctuations, we select events with a fixed 
number of wounded nucleons $n$, corresponding to the most probable value for this choice of $\sigma_{NN}^{\rm inel}$ 
(this value is obtained by first running a minimum-bias calculation and then choosing the most frequent value of $n$).
In order to isolate the effect of the twin correlations, we switch off the nuclear correlations studied in 
Sec.~\ref{sec:NNcorrelations} as well as  the correlations from recentering (cf. Appendix~\ref{sec:recenter}): 
as a consequence, $g_A(\x,\y)$ in Eq.~(\ref{eq:twin}) is a constant
(see Eq.~(\ref{eq:gnocorr})). 
The half-integrated pair distribution function $g(\Delta)$, with $\Delta=\x-\y$, clearly shows the sharp positive peak expected from Eq.~(\ref{eq:twin}). The finite width corresponds to the finite value of $\sigma_{NN}^{\rm inel}$.
Note that the asymptotic value at large relative distance $\Delta$,
where correlations are negligible, is 
slightly smaller than unity, and approximately given by
Eq.~(\ref{eq:gnocorr}). 

\begin{figure}[tb]
\begin{center}
\includegraphics[angle=0,width=0.5 \textwidth]{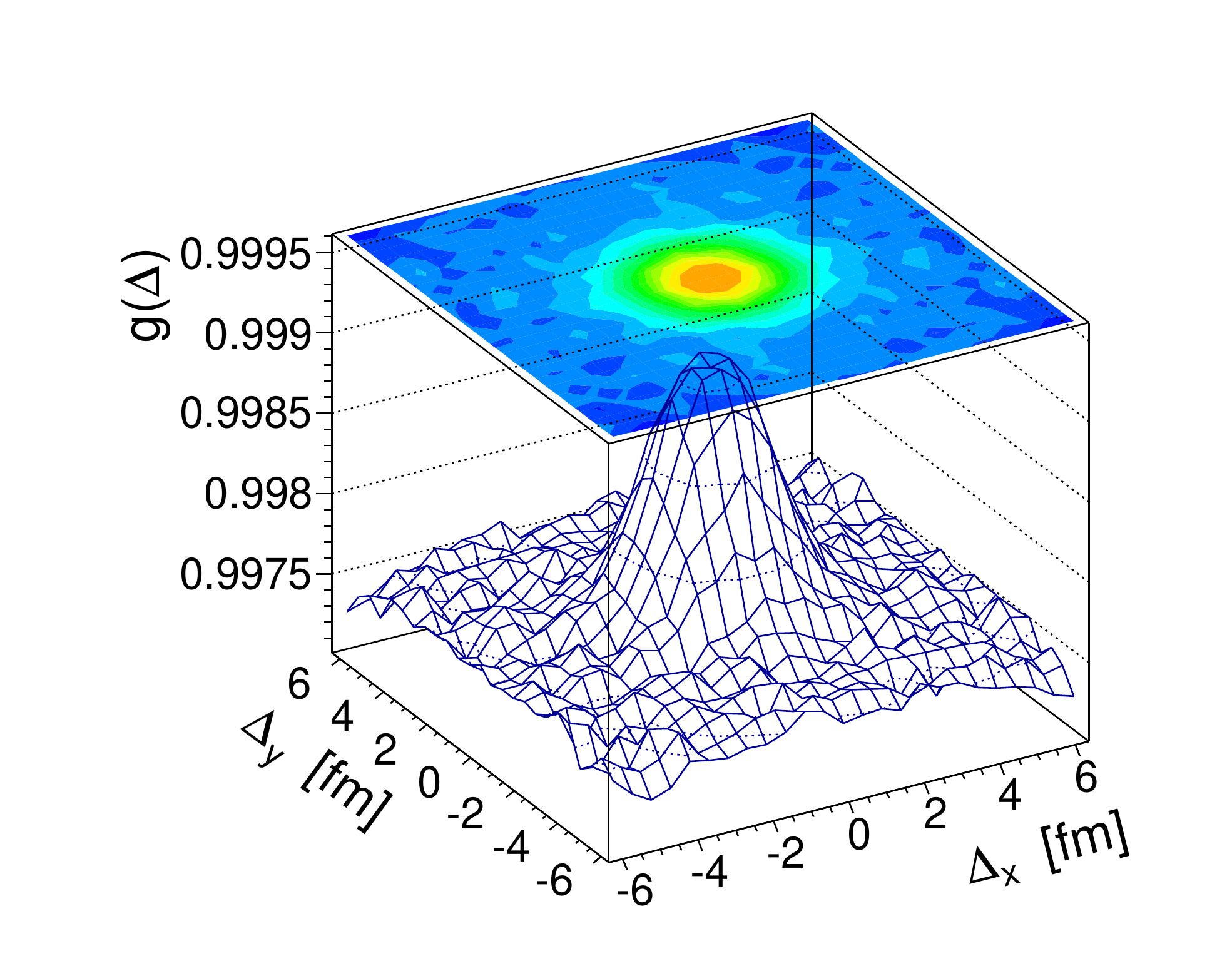} 
\end{center}
\vspace{-2mm}
\caption{(Color online) Same as Fig.~\ref{fig:f20} but for
with $\sigma_{NN}^{\rm inel}=68$~mb and $n=410$ (i.e., we again fix $n$ to its most probable value). 
We observe a significant melting of the peak compared to Fig.~\ref{fig:f20}.
\label{fig:f68}
}
\end{figure}
Fig.~\ref{fig:f68} illustrates the effect of increasing the cross section up to the actual value at the LHC, $\sigma_{NN}^{\rm inel}=68$~mb.  Naturally, the width of the correlation peak increases proportionally to $\sqrt{\sigma_{NN}^{\rm inel}}$. 
Meanwhile, the height of the peak decreases by a factor $\sim 35$, such that the
integral of the peak decreases by a factor $\sim 10$. We have checked that as
one further increases  $\sigma_{NN}^{\rm inel}$, the peak broadens and
completely melts down:
The twin correlations disappear. 
This can also be easily understood:  In the limit of infinite $\sigma_{NN}^{\rm
inel}$, {\it all\/} the nucleons are wounded, hence become uncorrelated. 

The dimensionless control parameter separating the regimes of
``small'' and ``large'' wounding cross section is the 
average number of target nucleons hit by each projectile nucleon,
which we denote by $\cal N$: this is roughly the product of the average
density $A/(\pi R_A^2)$ with $R_A$, the nuclear radius, and the cross section
$\sigma_{NN}^{\rm inel}$. Taking $R_A\simeq A^{1/3}r_0$, with
$r_0=1.2$~fm, one obtains
\begin{equation}
{\cal N}\sim \frac{A^{1/3}\sigma_{NN}^{\rm inel}}{\pi r_0^2}\simeq
4\frac{\sigma_{NN}^{\rm inel}[{\rm fm}^2]}{\pi},
\label{eq:control}
\end{equation}
where, in the last equality, we have chosen $A=208$. 
The value $\sigma_{NN}^{\rm inel}=20$~mb chosen in 
Fig.~\ref{fig:f20} corresponds to ${\cal N}\sim 2.6$, a number
significantly larger than unity: The twin correlations are already
significantly suppressed for this value of the cross section, which
explains why the peak is only a few percent above unity. 

The strength of the twin correlations  can also be investigated differentially as a function of the position in the transverse plane. 
For realistic values of $\sigma_{NN}^{\rm inel}$, all nucleons in the center are wounded, and the twin correlations are only present 
near the boundary of the fireball. This expectation is confirmed by our numerical simulations. 

Finally, if one takes into account repulsive nuclear correlations (Sec.~\ref{sec:NNcorrelations}), the correlation functions of Figs.~\ref{fig:f20} and \ref{fig:f68} display an additional central dip, as in Fig.~\ref{fig:f1}~(b), thus partially hiding the effect of the twin correlations. 

\subsection{Fluctuation observables}

\begin{figure}[tb]
\begin{center}
\includegraphics[angle=0,width=0.45 \textwidth]{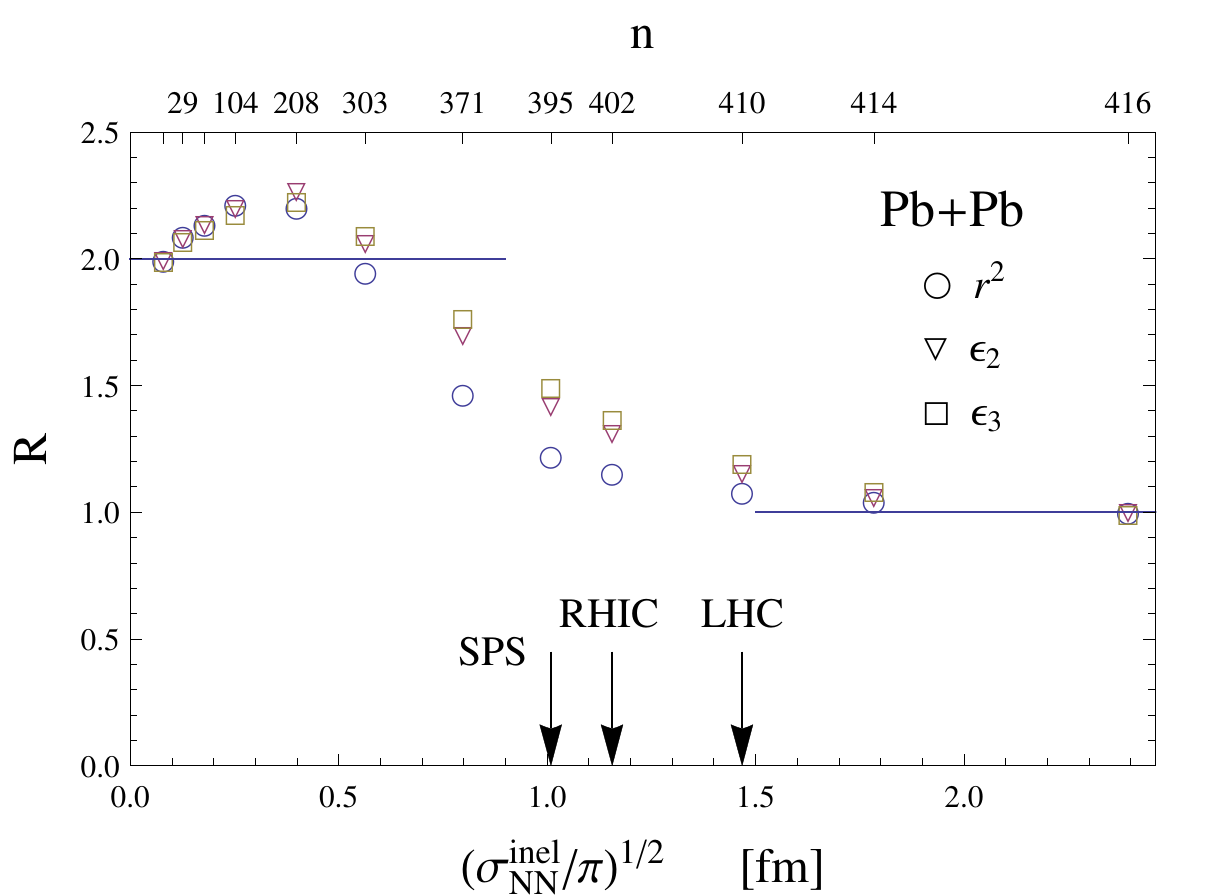} 
\end{center}
\vspace{-3mm}
\caption{(Color online) The ratios defined in Eq.~(\ref{eq:ratioR}) for the Pb+Pb collisions at various values 
of the total inelastic NN cross section, $\sigma_{NN}^{\rm inel}$. The corresponding fixed values of the number 
of the wounded nucleons, $n$, is shown on the upper horizontal axis.
\label{fig:PbPb_ns}}
\end{figure}   
We now study  numerically the effect of the twin correlations on fluctuation observables, namely, the variances of 
$r_{\rm rms}^2$, $\varepsilon_2$, $\varepsilon_3$. 
To this end, we evaluate for each observable the ratio $R$ in Eq.~(\ref{eq:ratioR}), which gives its relative increase due to correlations. 
These ratios are plotted in Fig.~\ref{fig:PbPb_ns} as a function of $\sigma_{NN}^{\rm inel}$.\footnote{We vary $\sigma_{NN}^{\rm inel}$ by varying the wounding radius $\sigma_w$ in Eq.~(\ref{eq:parth}), while $A$ is kept constant.}
As in Sec.~\ref{sec:twin}, we simulate central Pb-Pb collisions, where nuclear correlations are switched off, and we fix the number of wounded nucleons $n$ to its most probable value for each value of of $\sigma_{NN}^{\rm inel}$. 
The limits $\sigma_{NN}^{\rm inel} \to 0$ and $\sigma_{NN}^{\rm inel} \to \infty$ are readily understood from the discussion of Sec.~\ref{sec:twin}. 
For small $\sigma_{NN}^{\rm inel}$, the twin correlations double the
density-density correlation with respect to the uncorrelated case, hence all
ratios tend to 2. 
For large  $\sigma_{NN}^{\rm inel}$, the twin correlations vanish and all ratios
approach 1. 
The behavior between these two limits controlled by
the parameter ${\cal N}$ in Eq.~(\ref{eq:control}), which is 1 for 
$(\sigma_{NN}^{\rm inel}/\pi)^{1/2}\simeq 0.5$.
This behavior is nontrivial: in particular, all ratios increase above
2 before deceasing to 1. 
This increase is an effect of induced 
secondary correlations: a nucleon from nucleus $A$ wounds a nucleon from nucleus $B$, 
which in turn wounds another nucleon from nucleus $A$, thus inducing a correlation between participants 
of nucleus $A$. 
Note that while all three curves have the same asymptotic limits, intermediate values differ depending on the observable. 
At the values of the wounding cross section corresponding to the collisions at 
RHIC and the LHC (42~mb, and 68~mb, respectively), the ratios are closer to the
uncorrelated limit. 

Our result at the RHIC energy is somewhat smaller than that of Alver et
al.~\cite{Alver:2008zza}, 
where the ratios of physical to mixed-event results were presented.
However, their calculation is slightly different:
in particular, they do not fix the impact parameter or the number of
participants, thus including more 
sources of fluctuations which increase the variance.
Similarly,  it was found in Ref.~\cite{Bhalerao:2011bp}
that correlations increase the variance of $\varepsilon_2$ and $\varepsilon_3$ by a factor $\sim 2$ at RHIC and the LHC. 
The difference with our result is likely due to the fact that the present simulation
gives an identical weight to each wounded nucleon, in contrast with
the usual implementation at RHIC or the LHC where weights increase
linearly with the number of binary collisions~\cite{Miller:2007ri}.   
Generally, one expects that any additional source of fluctuations~\cite{Alvioli:2013vk,Rybczynski:2013mla}
will increase the variance, thus producing an effect similar to the twin
correlations. 

\begin{figure}[tb]
\begin{center}
\includegraphics[angle=0,width=0.45 \textwidth]{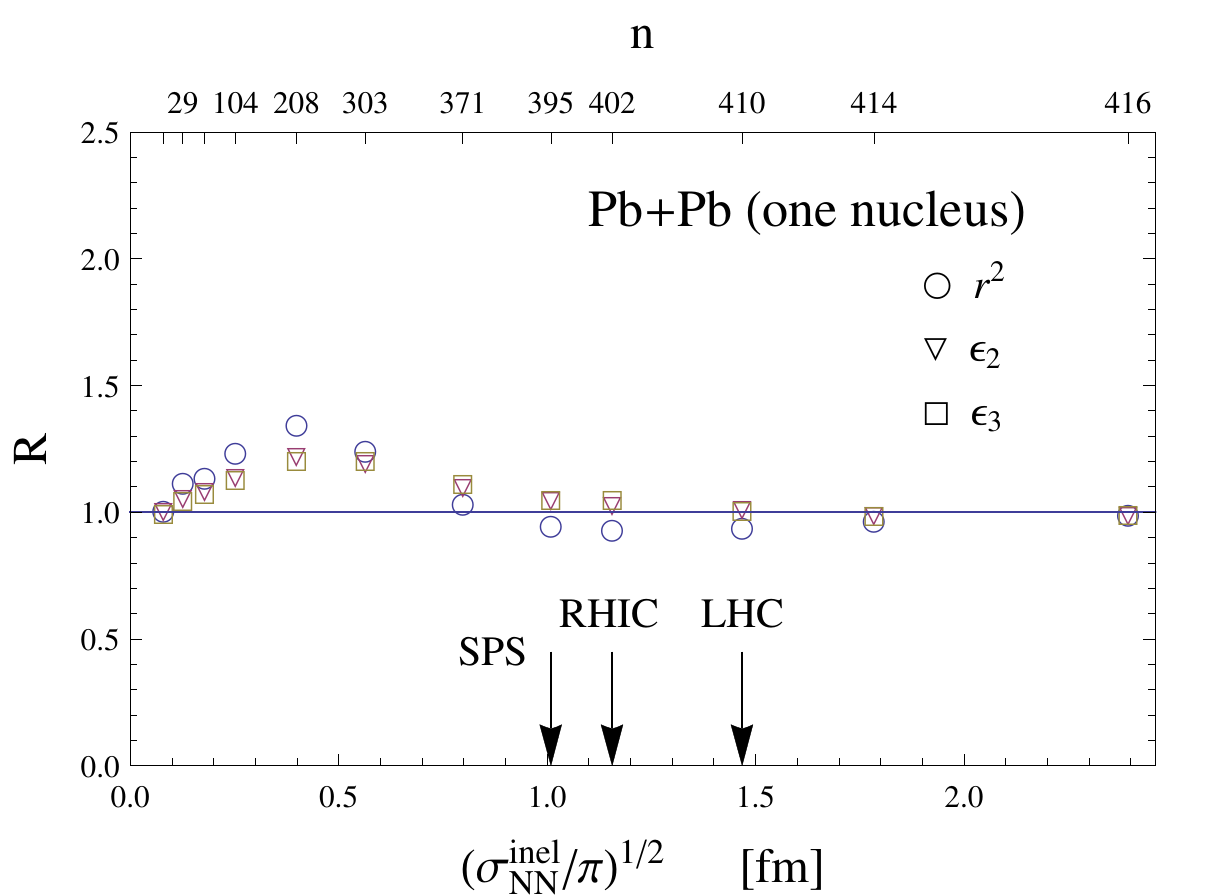} 
\end{center}
\vspace{-3mm}
\caption{(Color online) Same as Fig.~\ref{fig:PbPb_ns} but for the case where wounded nucleons coming only from one nucleus are taken into account.\label{fig:PbPb_A_ns}}
\end{figure}   
In order to further investigate the origin of correlations, in
particular confirm the mechanism of secondary correlations, 
we now repeat the simulation using a modified Glauber model, where
participants from only one nucleus contribute to the density. 
In other terms, we assume $\rho(\x)=\rho_A(\x)$, following the notations of Sec.~\ref{sec:twin}. 
In the wounded nucleon model, this corresponds to the density at very forward rapidity~\cite{Bozek:2010vz}. 
This modification effectively switches off direct twin correlations, which 
involve nucleons from different nuclei. 
The resulting values of $R$ are displayed in Fig.~\ref{fig:PbPb_A_ns}.
Correlations now disappear in both limits of large and small $\sigma_{NN}^{\rm inel}$, as expected. 
The departure from unity at intermediate values of $\sigma_{NN}^{\rm
  inel}$ is again an effect of induced 
secondary correlations. 
We note that at the SPS, RHIC, and LHC energies, the secondary correlations
are negligible.

\begin{figure}[tb]
\begin{center}
\includegraphics[angle=0,width=0.45 \textwidth]{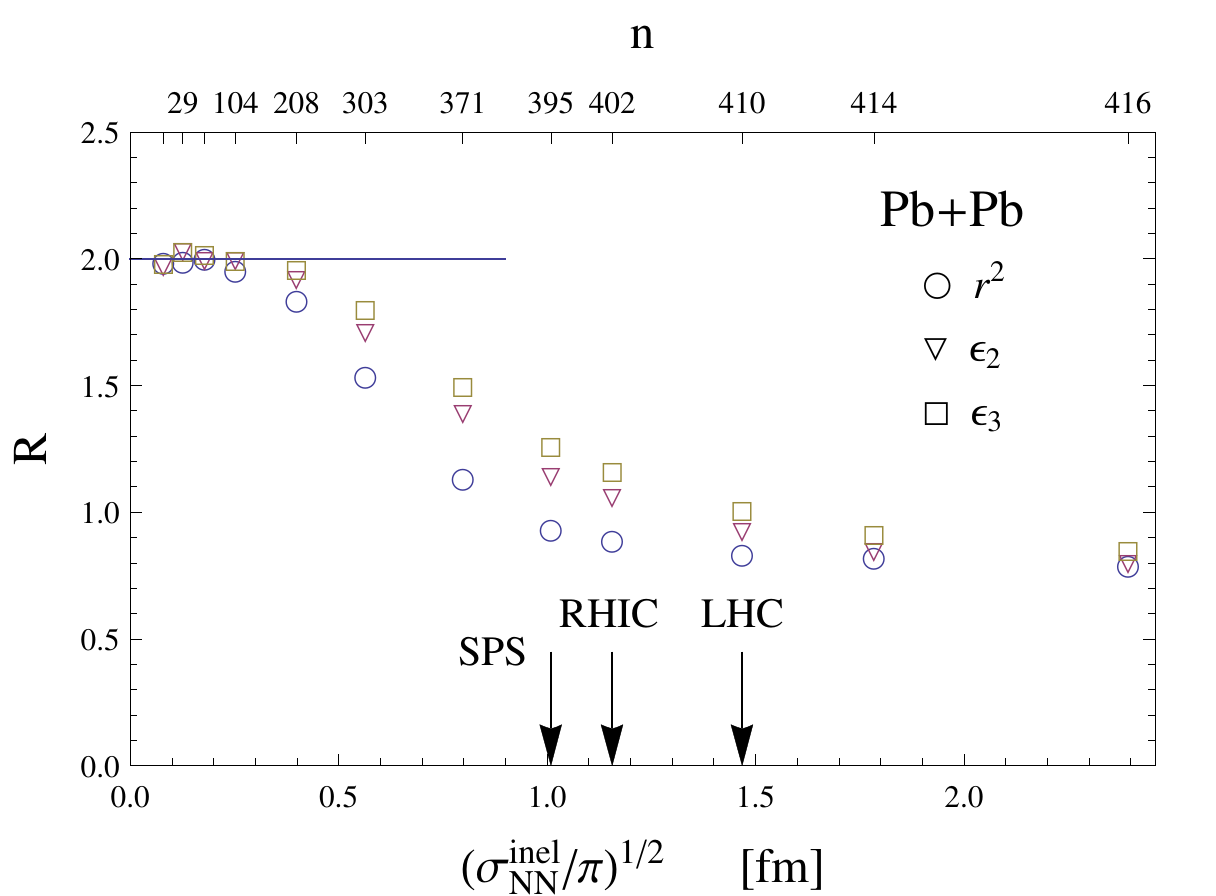} 
\end{center}
\vspace{-3mm}
\caption{(Color online) Same as Fig.~\ref{fig:PbPb_ns} but for the case where nuclear distributions of Ref.~\cite{Alvioli:2009ab} with NN repulsion are used.
\label{fig:PbPb_ns_d}}
\end{figure}  
Finally, we study the combined effects of the twin correlations and nuclear correlations~\cite{Alvioli:2009ab}. 
Results are presented in Fig.~\ref{fig:PbPb_ns_d}. 
As already shown in Ref.~\cite{Broniowski:2010jd}, repulsive correlations result in a sizable decrease of fluctuation 
observables, which is at the level of 20\%  for large $\sigma_{NN}^{\rm inel}$. 
Let us analyze the origin of this result. Much of the discussion of
Sec.~\ref{sec:analytic} can be carried over to the nucleus-nucleus
collisions. For large $\sigma_{NN}^{\rm inel}$, the twin correlations are
negligible, and the only correlations among sources are those already
present in the colliding nuclei. We therefore write
\begin{equation}\label{eq:f2NN}
f^{(2)}(\x,\y)\simeq f^{(1)}(\x)f^{(1)}(\y)
\left(1-\frac{1}{2}d(|\x-\y|)\right),
\end{equation}
where the factor $1/2$ accounts for the fact that only pairs of
sources from the same nucleus are correlated (neglecting $1/n$
corrections). 
In writing Eq.~(\ref{eq:f2NN}), we  neglect the small change of
normalization induced by the repulsive correlation, i.e., we assume
\begin{equation}
\int \rmd^2\x \rmd^2\y f^{(1)}(\x)f^{(1)}(\y)d(|\x-\y|)\ll 1. 
\end{equation}
Inserting Eq.~(\ref{eq:f2NN}) into Eq.~(\ref{eq:gconstantn}), 
and ignoring small corrections of order $1/n$, one obtains
\begin{equation}
\label{eq:gxynotwin}
g(\x,\y)\simeq 1-\frac{1}{2}d(|\x-\y|).
\end{equation} 
Inserting this equation into Eq.~(\ref{eq:ssxy}) yields
\begin{eqnarray}
\label{eq:ssxynotwin}
S(\x,\y)&\simeq&\langle\rho(\x)\rangle\delta(\x-\y)-\frac{1}{2}\langle\rho(\x)\rangle
\langle\rho(\y)\rangle d(\x-\y)\cr
&\simeq&\langle\rho(\x)\rangle\delta(\x-\y)-\frac{1}{2}\langle\rho(\x)\rangle^2
d(\x-\y),
\end{eqnarray}
where in the second line we have used the fact that the
range of the nucleon-nucleon correlation is much shorter than the
nuclear radius, such that $\rho(\x)\simeq\rho(\y)$.  
Note that Eq.~(\ref{eq:ssxynotwin}) does not satisfy the normalization
condition Eq.~(\ref{eq:Snorm}): the integral of $S(\x,\y)$ should be 0. A more
careful calculation (enforcing the proper normalization in 
Eq.~(\ref{eq:f2NN}) and restoring terms of order $1/n$ in
Eq.~(\ref{eq:gxynotwin})) shows that 
$S(\x,\y)$ has an additional disconnected term of the form 
$c\langle\rho(\x)\rangle\langle\rho(\y)\rangle$. 
This disconnected term, however, does not contribute to
the $\varepsilon_n$ fluctuations for symmetry reasons, as already noted at
the end of Sec.~\ref{sec:res}. 

Inserting Eq.~(\ref{eq:ssxynotwin}) into Eqs.~(\ref{eq:variances}) and
(\ref{eq:ratioR}) results in
\begin{eqnarray}
R(\varepsilon_l)&=&1-\frac{1}{2}\frac{\int\langle\rho(\x)\rangle^2 r^{2l}
  \rmd^2\x}
{\int\langle\rho(\x)\rangle r^{2l} \rmd^2\x}\int d(s)\rmd^2\s. 
\end{eqnarray}
To get numerical estimates we take a rough approximation where
$\langle\rho(\x)\rangle$ is a projection of a uniform sphere of radius $R_A$ on
a plane, $\langle\rho(\x)\rangle = 6 n \sqrt{R_A^2-x^2}/(4 \pi R_A^3)$. Then
\begin{eqnarray}
R(\varepsilon_l)&\sim& 1-\frac{3 n B \sigma_d^2 \Gamma(l+5/2)}{\sqrt{\pi}
R_A^2 \Gamma(l+3)}, 
\end{eqnarray}
which yields for $n=416$ the values $R(\varepsilon_2)\sim 0.71$ and
$R(\varepsilon_3)\sim 0.74$, in approximate agreement (at the level of 10\%)
with the detailed
simulation
shown in Fig.~\ref{fig:PbPb_ns_d}.

As in the case of the proton-nucleus collisions (Eq.~(\ref{eq:R4})), the
effect of nuclear correlations is enhanced by a factor $n$. 
Our results show that a repulsive short-range correlation always
reduces the $\varepsilon_n$ fluctuations. 
It has been found numerically~\cite{Alvioli:2011sk} that when 
realistic two- and three-body correlations are implemented, the rms
anisotropy barely changes, which suggests that the short-range
repulsive NN correlation is compensated by a intermediate-range attractive
NN correlation. 

\section{Conclusions}

We have shown that quantities characterizing fluctuations in the early phase of
relativistic heavy-ion and proton-nucleus collisions --- specifically, the size
and eccentricity fluctuations ---  can be generally expressed in terms of the
density-density correlation. 
We have analyzed the structure of this density-density correlation
within the Monte Carlo Glauber model. 
It can be written as the sum of an autocorrelation part, which is the
contribution of fluctuations, and a part involving genuine two-body
correlations.  

We have investigated in detail the effect of these
genuine correlations on selected observables. 
We have shown that effects of the nucleon-nucleon correlations inside
colliding nuclei are parametrically enhanced by a factor $n$, where
$n$ is the number of participants, yet they are expected to be small
with realistic interactions~\cite{Alvioli:2011sk}. 
For nucleus-nucleus collisions, we have identified a new type of
correlations, due to the collision mechanism itself,  
which we have dubbed twin correlations; they increase fluctuations. 
At the RHIC and LHC energies, however, the nucleon-nucleon cross section
is so large that essentially all nucleons in the interaction region
are wounded. As a result, the twin correlations are small and localized
near the boundary of the fireball. 

All sources of correlations studied in this paper --- namely,
the autocorrelation, nuclear correlations, and the twin correlations --- 
involve scales much shorter than the nuclear radius. The eccentricity
and size fluctuations in the Glauber model appear then to be
created by uncorrelated, small-scale fluctuations in the transverse plane. 
Subnucleonic fluctuations, which are not incorporated in the Glauber model, are
also intrinsically small-scale phenomena. They typically increase the magnitude
of the local fluctuations, but do not give rise to any large-distance
correlation. To a good approximation, the Monte-Carlo Glauber model provides a
picture of energy deposition for the RHIC at LHC energies in terms of
independent sources, that seems to capture the main features of these
fluctuations and their correlations. 

\begin{acknowledgments}
This work was supported by the European Research Council under the
Advanced Investigator Grant ERC-AD-267258, and by the Polish National Science
Centre, grants DEC-2012/06/A/ST2/00390 and DEC-2011/01/D/ST2/00772.  
JYO thanks Derek Teaney for discussions and the LNS at MIT for hospitality. 
\end{acknowledgments}

\appendix

\section{Distributions of sources \label{sec:cps}}

The sources are defined by their locations $\{\x_i\}$ in the transverse plane
and by their number $n$. The positions $\{\x_i\}$ and the number $n$ are random
variables that fluctuate from event to event. 
For a given $n$, we denote by $f_n(\x_1, ..., \x_n)$ the probability
distribution that sources are localized at $\x_1,..., \x_n$. 
By definition, $f_n$ is completely symmetric. 
The
$k$-body distribution is defined by integrating over $n-k$ positions:  
\begin{equation}
f_n^{(k)}(\x_1,..,\x_k)=\int \rmd^2 \x_{k+1}... \rmd^2\x_n\, f_n(\x_1, ..., \x_n). \label{eq:marg}
\end{equation} 
In particular, the one-body density can be obtained by integrating the 2-body
density: 
\begin{equation}
f_n^{(1)}(\x_1)=\int \rmd^2\x_2 \,f_n^{(2)}(\x_1,\x_2). \label{eq:f1f2}
\end{equation}
Note that in general the  $k$-body distribution thus defined depends
on the total number of sources $n$.
Note also that $\int\rmd^2 x\,f_n^{(1)}(\x)=1$.

The average value of the density $\rho(\x)$ (Eq.(\ref{eq:deltasources})) is 
\begin{eqnarray}
 \langle \rho(\x) \rangle &=& \langle \int  \rmd^2\x_1... \rmd^2\x_n \,f_n(\x_1, ..., \x_n) \sum_{i=1}^n \delta(\x-\x_i) \rangle \cr
&=& \langle n f_n^{(1)}(\x) \rangle, 
\label{eq:eps}
\end{eqnarray}
where we have allowed the multiplicity $n$ to fluctuate for sake of generality, and 
the angular brackets in the right-hand side of Eq.~(\ref{eq:eps}) denote an average over the distribution of $n$.
Similarly,
\begin{eqnarray}
&& \langle \rho(\x) \rho(\y) \rangle = \nonumber \\
&& \langle \int  \rmd^2\x_1... \rmd^2\x_n \,f_n(\x_1, ..., \x_n)\sum_{i,j}  \delta(\x\!-\!\x_i)\delta(\y\!-\!\x_j)\rangle  \nonumber \\
&& = \langle n f_n^{(1)}(\x) \rangle \delta(\x-\y) + \langle n(n-1) f_n^{(2)}(\x,\y) \rangle \label{eq:ee} .
\end{eqnarray}
Equations (\ref{eq:eps},\ref{eq:ee}) yield the following generic
decomposition of the density-density correlation function (\ref{eq:defcor}):
\begin{eqnarray}
S(\x,\y)&=& \langle n f_n^{(1)}(\x) \rangle \delta(\x-\y) \label{eq:sxy}  \\ &+& \langle n(n-1) f_n^{(2)}(\x,\y) \rangle  
-\langle n f_n^{(1)}(\x) \rangle \langle n f_n^{(1)}(\y) \rangle. \nonumber
\end{eqnarray}
The first term is commonly referred to as an {\em autocorrelation}, the second term is the {\em inclusive} 
distribution of pairs, and the last term is the {\em disconnected} part, 
composed of the product of the inclusive single-particle distributions.
Note that the autocorrelation term follows from the rearrangement of the sums in the definition  Eq.~(\ref{eq:ee}). It represents a major contribution to the correlation function and is by all means physical.
Using Eq.~(\ref{eq:eps}), one can rewrite Eq.~(\ref{eq:sxy}) as
\begin{eqnarray}\label{eq:sxy1}
S(\x,\y)&=& \langle \rho(\x) \rangle \delta(\x-\y) \nonumber \\ &+& \langle n(n-1) f_n^{(2)}(\x,\y) \rangle  
-\langle \rho(\x) \rangle \langle \rho(\y) \rangle. 
\end{eqnarray}
The {\em pair distribution function} is defined by
\begin{eqnarray}
g(\x,\y) =\frac{\langle n(n-1) f_n^{(2)}(\x,\y) \rangle}{ \langle \rho(\x) \rangle \langle \rho(\y) \rangle } . \label{eq:gdef}
\end{eqnarray}
Inserting into Eq.~(\ref{eq:sxy1}), one obtains Eq.~(\ref{eq:ssxy}). 
Occasionally, we also use a differently normalized pair correlation function 
\begin{eqnarray}
 P(\x,\y)=\frac{\langle n \rangle^2}{\langle n(n-1) \rangle} g(\x,\y), \label{def:P}
\end{eqnarray}
which asymptotes to unity in the absence of correlations.

In this paper, we carry out simulations where $n$ is fixed. In this case, one
can drop the subscript $n$ and Eq.~(\ref{eq:gdef}) 
simplifies to: 
\begin{equation}
\label{eq:gconstantn}
g(\x,\y) =\left(1-\frac{1}{n}\right)
\frac{f^{(2)}(\x,\y)}{f^{(1)}(\x)f^{(1)}(\y)}.  
\end{equation}
When no correlations are present, the $n$-particle distribution function is a product of the single-particle distributions. In particular,
\begin{eqnarray}
f^{(2)}(\x,\y)=f^{(1)}(\x)f^{(1)}(\y), 
\end{eqnarray}
and $g(\x,\y)$  takes the form displayed in Eq.~(\ref{eq:gnocorr}), while $P(\x,\y)=1$.

\section{Superposition models \label{app:sup}}

In Eq.~(\ref{eq:deltasources}), we assume that the strength of each source is the same. 
This restriction is lifted in {\em superposition models}, where the
strength of the source is allowed to fluctuate. 
In this case, one introduces an additional random variable $w_i$ that measures the strength of the source $i$,  
and writes the density $\rho(\x)$ as
\begin{eqnarray}
\rho(\x)=\sum_{i=1}^n w_i \delta(\x-\x_i), \label{eq:w} 
\end{eqnarray}
with the weights $w_i$ generated according to some suitable statistical distribution. 

One generally assumes for simplicity that the weight of a source, $w_i$, is uncorrelated with its location 
$\x_i$. One also  assumes that weights of different sources are not correlated with one another 
or with the multiplicity $n$. 
Then, Eqs.~(\ref{eq:eps}) and (\ref{eq:ssxy}) are replaced
respectively by 
\begin{eqnarray}\label{eq:epssup}
 \langle \rho(\x) \rangle &=& 
\langle w\rangle \langle n f_n^{(1)}(x) \rangle, 
\end{eqnarray}
and
\begin{eqnarray}
S(\x,\y) &=&  \frac{\langle w^2\rangle}{\langle w\rangle}
\langle \rho(\x) \rangle \delta(\x-\y) \nonumber \\ &+&
\langle \rho(\x) \rangle \langle \rho(\y) \rangle [g(\x,\y)-1], \label{eq:supmodel}
\end{eqnarray}
while the pair distribution function remains given by Eq.~(\ref{eq:gdef}).
Comparison of Eq.~(\ref{eq:supmodel}) with Eq.~(\ref{eq:ssxy}) shows  
that fluctuations of the weight $w$ enhance the relative contribution of the autocorrelation term. 

\section{Smearing \label{sec:smear}}

In a more realistic situation, the sources may be attributed a finite width. 
This can be implemented by smearing the point-like source with a finite-width function, $s$: One replaces 
Eq.~(\ref{eq:w}) by  
\begin{eqnarray}
\rho(\x)=\sum_{i=1}^n w_i s(\x-\x_i).
\end{eqnarray}
Then, Eqs.~(\ref{eq:epssup}) and (\ref{eq:ee}) are replaced by
\begin{eqnarray}
\langle \rho(\x) \rangle &=&\langle w \rangle
\int \rmd^2\x_1 s(\x-\x_1) \langle n f_n^{(1)}(\x_1) \rangle,
\end{eqnarray}
and
\begin{eqnarray}
&& \langle \rho(\x) \rho(\y) \rangle = \nonumber \\
&& \langle w^2 \rangle
\int  \rmd^2\x_1 s(\x_1-\x) s(\x_1-\y) \langle n  f_n^{(1)}(\x_1) \rangle + \nonumber \\
&& \langle w \rangle^2 \int  \rmd^2\x_1 \rmd^2\x_2 s(\x_1-\x) s(\x_2-\y) \langle n  f_n^{(2)}(\x_1,\x_2) \rangle.\nonumber\\
\end{eqnarray}
Note that the autocorrelation term is no longer singular. A typical choice for the smearing function
in practical applications is a Gaussian of a width of a fraction of a
fermi~\cite{Holopainen:2010gz,Schenke:2010rr,Bozek:2012gr}.

\section{Correlations from recentering \label{sec:recenter}}

For completeness, we also discuss the long-range correlations present in the
Monte Carlo 
simulations due to {\em recentering}, i.e., the condition that the center of mass of each nucleus
is fixed at a given location in each event. 
Recentering is implemented in calculating initial anisotropies: they
are 
evaluated in a coordinate system where the fireball is centered~\cite{Teaney:2010vd}.

In each event, the two-dimensional transverse coordinates
$\x_1,\cdots,\x_n$ are independent. We denote their distribution by 
$f(\x_i)$. Without loss of generality we may assume 
$\langle \x_i \rangle =\int \x_i f(\x_i)\rmd\x_i=0$.
The recentered coordinate is defined by 
\begin{eqnarray}
\label{recentering}
&&\x'_1
=\x_1-\frac{\x_1+\cdots+\x_n}{n}
=\frac{n-1}{n}\left(\x_1-\frac{\x_2+\cdots+\x_n}{n-1}\right) \nonumber \\
&&=\frac{n-1}{n}\left(\x_1-\T\right),
\end{eqnarray}
where we have introduced the auxiliary variable
\begin{equation}
\T\equiv\frac{\x_2+\cdots+\x_n}{n-1}.
\end{equation}
In the limit $n\gg 1$, the distribution of $\T$ is Gaussian according
to the central limit theorem. Its normalized distribution is $P_{n-1}(\T)$, where
\begin{equation}
P_k(\T)={\frac{k}{2\pi\sigma^2}}\exp\left(-\frac{k T^2}{2\sigma^2}\right),
\end{equation}
with $\sigma^2\equiv\langle \x_i^2\rangle$ assumed to be identical
for all $i$. 
Furthermore, $\x_1$ and $\T$ are independent variables, such that the
distribution of $\x'_1$ as defined in Eq.~(\ref{recentering}) is
\begin{eqnarray}
f_n^{(1)}(\x'_1)
&=&\int f(\x_1) \rmd^2\x_1 P_{n-1}(\T)\rmd^2\T\cr
&&\times \delta\left(\x'_1-\frac{n-1}{n}(\x_1-T)\right) \cr
&=&\frac{n}{n-1}\int f\left(\frac{n\x'_1}{n-1}+\T\right)P_{n-1}(T)\rmd^2\T.
\nonumber \\
\end{eqnarray}
Then we find easily
\begin{eqnarray}
&&\langle \x'_1 \rangle_s = \langle \x_1 \rangle \; (=0), \; \nonumber \\
&& {\rm Var}_s(\x'_1)=\left ( 1- \frac{1}{n^2}\right ) \sigma^2,
\end{eqnarray}
Hence the recentered distributions are shrunk by a term of the order $1/n^2$.

Carrying out a similar calculation for the case of the two-particle distributions we arrive at
\begin{eqnarray}
f_n^{(2)}(\x'_1,\x'_2)
&=&\int f(\x_1) \rmd^2\x_1 f(\x_2) \rmd^2\x_2 P_{n-2}(\T)\rmd^2\T \times
\nonumber  \\
&& \delta\left(\x'_1-\x_1+\frac{\x_1+\x_2}{n}+\frac{n-2}{n} \T \right) \nonumber \times \\
&& \delta\left(\x'_2-\x_2+\frac{\x_1+\x_2}{n}+\frac{n-2}{n} \T \right),
\nonumber \\
\end{eqnarray}
which yields
\begin{eqnarray}
 {\rm cov}(\x'_1,\x'_2)=-\frac{2\sigma^2}{n^2}.
\end{eqnarray}
Thus the correlations from recentering are small if the number of
sources $n$ is large.

\section{Distribution of participants in p-A collisions\label{sec:pAanalytic}}

Transverse positions of nucleons within the target nucleus at the time
of impact are random variables.  
The probability distribution that the $A$ nucleons in the nucleus are in a configuration 
$(\x_1,\dots,\x_A)$ is projected on the transverse plane, yielding the distribution 
$T(\s_1, \dots, \s_A)$, where $\s_i$ denotes the position of the $i$th nucleon in the transverse plane.
By construction, $\int \rmd\s_1\dots \rmd\s_A T(\s_1, \dots, \s_A)=1$. 
In the absence of correlations, the function $T(\s_1, \dots, \s_A)$ is
of the form $T(\s_1, \dots, \s_A)= T_0(\s_1)\dots T_0(\s_A)$.  In
order to take into account the short range two-body correlations in
the nucleus wave function 
(Sec.~\ref{sec:NNcorrelations}), we use the simple ansatz  
\begin{equation}
\label{eq:TA}
T(\s_1, \dots,\s_A)=c T_0(\s_1)\dots T_0(\s_A) \prod_{\stackrel{i,j=1}{i<j}}^A (1-d(|\s_i-\s_j|)), 
\end{equation}
where $c$ is a normalization constant.

The probability  
that the proton, incident at an impact parameter $\b$, interacts inelastically with $n$ selected participants  in a given 
configuration, and does not interact with the remaining $A-n$ nucleons  (called spectators), also in a given configuration, is proportional to 
\begin{eqnarray}
&& \theta(\b-\s_1) \dots \theta(\b-\s_n) \times \label{eq:glgen} \\ 
&& ~~(1-\theta(\b-\s_{n+1}))\dots (1-\theta(\b-\s_{A})) T(\s_1, \dots, \s_A), \nonumber
\end{eqnarray}
where $\theta(s)$ is the wounding profile, Eq.~(\ref{eq:parth}). 
By integrating over the coordinates of the spectators, one obtains the probability  $p_n(\s_1,\dots \s_n;\b) $ to find $n$ participants at positions $\{\s_1,\cdots, \s_n\}$. 

We now derive Eq.~(\ref{eq:f2gl}). For a given number of participants
$n$, the two-body distribution of participants is obtained by
integrating Eq.~(\ref{eq:glgen}) over $\s_3,\cdots,\s_A$. 
The terms which do not depend on $\s_3,\cdots,\s_A$ are 
\begin{equation}
\label{eq:f2gl1}
\theta(\b-\s_1)\theta(\b-\s_2)T_0(\b-\s_1)T_0(\b-\s_2)(1-d(|\s_1-\s_2|).
\end{equation}
The remaining terms depend on $\s_1$ and $\s_2$ only through nuclear
correlations. 
In order to estimate the magnitude of these terms, 
we use the fact  that $d$ is  small, as can be seen, for instance, 
in Fig.~\ref{fig:arho}, where numerically we have
$d(\rho)\lesssim 0.15$. 
Expanding to first order in $d(|\s_i-\s_j|)$, 
one rewrites the correlation term in Eq.~(\ref{eq:TA}) as 
\begin{eqnarray}
&&\prod_{\stackrel{i,j=1}{i<j}}^A (1-d(|\s_i-\s_j|))=(1-d(|\s_1-\s_2|))\cr
&&\times 
(1-\sum_{i\ge 3}d(|\s_1-\s_i|)-\sum_{i\ge 3}d(|\s_2-\s_i|))\cr
&&\times \prod_{\stackrel{i,j=3}{i<j}}^A (1-d(|\s_i-\s_j|)).
\end{eqnarray}
Upon integration over $\s_3,\cdots,\s_A$, the terms in the second line
of the right-hand side can be written in the form 
$k(1-\epsilon(\s_1)-\epsilon(\s_2))\simeq
k(1-\epsilon(\s_1))(1-\epsilon(\s_2))$, where $\epsilon(\s_{1,2})$ is a
first-order correction proportional to $d$, and $k$ is independent of
$\s_1$ and $\s_2$. 
To first order, the effect of interactions is therefore to slightly
change the factors depending on $\s_1$ and $\s_2$ in Eq.~(\ref{eq:f2gl1}). This
modification can be neglected in a first approximation.

Since the range of nucleon-nucleon collisions, $\sigma_w$ in
Eq.~(\ref{eq:parth}), is much  
smaller than the nuclear radius, one can further neglect the variation
of the thickness function $T_0(\s)$ in Eq.~(\ref{eq:f2gl1}), which
then reduces to Eq.~(\ref{eq:f2gl}). This is a very good approximation
for central collisions ($b=0$) considered in Sec.~\ref{sec:anmod},   
where $T_0(\s)$ is close to its maximum $T_0(0)$. 

\bibliography{ref,hydr}

\end{document}